\documentclass{aa}
\usepackage{graphicx}
\usepackage{rotating}
\usepackage{latexsym}
\usepackage{amssymb}
\usepackage{natbib}
\usepackage{txfonts}

\newcommand{\HI}{\ion{H}{i}}
\newcommand{\HII}{\ion{H}{ii}}

\newcommand{\HeII}{\ion{He}{ii}}
\newcommand{\ArII}{[\ion{Ar}{ii}]}
\newcommand{\ArIII}{[\ion{Ar}{iii}]}

\newcommand{\SIII}{[\ion{S}{iii}]}
\newcommand{\SIV}{[\ion{S}{iv}]}
\newcommand{\NeII}{[\ion{Ne}{ii}]}
\newcommand{\NeIII}{[\ion{Ne}{iii}]}
\newcommand{\OIV}{[\ion{O}{iv}]}

\newcommand{\tel}{$T_{\rm e}$}
\newcommand{\den}{$n_{\rm e}$}
\newcommand{\micron}{$\mu$m}
\newcommand{\cm}{cm$^{-3}$}
\newcommand{\mup}{$M_{\rm up}$}
\newcommand{\msun}{$M_{odot}$}

\def\msun{\ifmmode M_{\odot} \else $M_{\odot}$\fi}
\def\zsun{\ifmmode Z_{\odot} \else $Z_{\odot}$\fi}
\def\lsun{\ifmmode L_{\odot} \else $L_{\odot}$\fi}
\def\mup{\ifmmode M_{\rm up} \else $M_{\rm up}$\fi}
\def\mlow{\ifmmode M_{\rm low} \else $M_{\rm low}$\fi}

\newcommand{\iizw}{II\,Zw\,40}
\newcommand{\henize}{He\,2--10}

\begin{document}
   \title{High spatial resolution mid-infrared spectroscopy of the starburst
          galaxies NGC\,3256, II\,Zw\,40 and Henize\,2-10\thanks{Based on
          observations obtained at the European Southern Observatory,
          La Silla, Chile (ID 70.B-0583).}}

   \author{N.\,L. Mart\'{\i}n-Hern\'{a}ndez\inst{1}
         \and
         D. Schaerer\inst{2,3}
          \and
         E. Peeters\inst{4}
	 \and
	 A.\,G.\,G.\,M. Tielens\inst{5}
	 \and
	 M. Sauvage\inst{6}
          }

   \offprints{N.L.\,Mart\'{\i}n-Hern\'{a}ndez,\\
             \email{Leticia.Martin@obs.unige.ch}}

   \institute{
              Instituto de Astrof\'{\i}sica de Canarias, V\'{\i}a Lactea,
              E-38200 La Laguna, Tenerife, Spain
        \and
              Observatoire de Gen\`{e}ve, 51 Chemin des Maillettes,
              CH-1290 Sauverny, Switzerland
        \and
              Laboratoire Astrophysique de Toulouse-Tarbes (UMR 5572),
              Observatoire Midi-Pyr\'en\'ees,
              14 Avenue E. Belin, F-31400 Toulouse, France
        \and
             NASA Ames Research Center, MS 245-6, Moffett Field, CA 94035, USA
        \and
	     SRON National Institute for Space Research and Kapteyn Institute,
	     P.O. Box 800, 9700 AV Groningen, Netherlands
	\and
              CEA/DSM/DAPNIA/SAp, CE Saclay, 91191 Gif sur Yvette
              Cedex, France
            }

   \date{}

\titlerunning{Mid-IR spectroscopy of starburst galaxies}
\authorrunning{N.\,L. Mart\'{\i}n-Hern\'{a}ndez et al.}

  \abstract
{}
{In order to show the importance of high
spatial resolution observations of extra-galactic sources when compared to
observations obtained with larger apertures such as ISO, we present $N$-band spectra (8--13~\micron) of some
locations in three starburst galaxies. In particular, the two
galactic nuclei of the spiral galaxy NGC\,3256, the compact IR
supernebula in the dwarf galaxy \iizw\ and the two brightest IR knots
in the central starburst of the WR galaxy \henize.}
{The spectra have been obtained with TIMMI2 on the ESO 3.6\,m
telescope. An inventory of the spectra in terms of atomic
fine-structure lines and molecular bands is presented.}
%The spectra show an ample variety in terms of continuum,
%lines and molecular band strength.
{
%The two nuclei of NGC\,3256 and the
%two IR knots in \henize\ show a rising dust continuum, PAH bands
%and a strong \NeII\ line. On the contrary, the IR knot in \iizw\ is
%characterised by a rather flat continuum, a strong \SIV\ line and does
%not show the presence of PAH bands.
%As in the case of our pilot work on NGC\,5253 \citep{martin:ngc5253},
We show
the great value of these high spatial resolution data at constraining properties such
as the extinction in the mid-IR, metallicity or stellar content (age,
IMF, etc.). Regarding this, we have constrained the stellar content of
the IR compact knot in \iizw\ by using the mid-IR fine-structure lines
and setting restrictions on the nebular geometry.

Considering the PAH bands,
we have constructed a new mid-/far-IR diagnostic diagram
%, adapted from that by \cite{peeters04},
based on the 11.2~\micron\ PAH and continuum, accessible to
ground-based observations.
We find that extra-galactic nuclei and star clusters observed
at high spatial resolution (as is the case of the TIMMI2 observations)
are closer in PAH/far-IR to compact \HII\ regions, while galaxies
observed by large apertures such as ISO are closer to exposed PDRs
such as Orion. This is certainly due to the aperture
difference, where the much larger ISO aperture likely includes much of the
surrounding PDRs while the
TIMMI2 slit measures mainly the central emission of the \HII\ region.

Finally, we find a dependence between the presence or
non-presence of PAHs and the hardness of the radiation field as
measured by the \SIV/\NeII\ ratio.  In particular, sources with PAH
emission have in general a \SIV/\NeII\ ratio $\lesssim 0.35$. We
investigate possible origins for this relation and conclude that
it does not necessarily
imply PAH destruction, but could also be explained by the PAH-dust
competition for FUV photons.
We have also considered the scenario where the low PAH
emission could just be a consequence of the relative contribution of
the different phases of the interstellar medium, in particular, the presence
of a pervasive and highly ionised medium.}
{}


   \maketitle
%
%________________________________________________________________

\section{Introduction}
\label{sect:intro}

Mid-infrared (MIR) observations have proved to be of great value at providing
spectral diagnostics to quantify massive star formation and to
distinguish between stellar (starburst) and other (AGN) activity
\citep[e.g.][]{genzel00}.  However, most of these diagnostics are based on
``integrated spectra'' which have been obtained through large
apertures e.g. with the {\it Infrared Space Observatory}, ISO, with
an aperture larger than 14\arcsec\ for ISO/SWS
\citep[cf.][]{schaerer99,thornley00,forster01,rigby04}.
Therefore, these
measurements are likely to consist of contributions from numerous
``knots'', the diffuse ISM, etc. rendering their interpretation
difficult. Furthermore, some diagnostics (e.g. relating PAH and
continuum emission, high and low excitation fine-structure lines) may
originate from different spatial regions and in consequence, it is by
no means clear if, and to what extent, such spatially integrated or
``global'' spectra can be used for various diagnostic purposes.

In this respect, it has already been demonstrated how the spatial
scale of the observations greatly determine the MIR appearance of
galaxies \citep[e.g.][]{martin:ngc5253,siebenmorgen04}. This is the
case of, for instance, the starburst galaxy NGC\,5253, where most of
the high-excitation fine-structure line fluxes measured by within ISO
aperture (e.g. \SIV) are emitted by a single knot of $\sim$ 0\farcs1,
while other sources within the aperture mainly contribute to the
low-excitation line fluxes. For instance, only $\sim$ 20\% of the
\NeII\ emission comes from the otherwise dominant compact knot
\citep{martin:ngc5253}.  This has important implications in the
interpretation of line fluxes in terms of properties of the stellar
content such as age, IMF, etc.

We have recently started gathering ground-based
high spatial resolution MIR observations of several nearby
starbursts using TIMMI2 on the 3.6\,m telescope, which provides a slit
width of 1\farcs2. Our pilot work on the well studied NGC\,5253
\citep{martin:ngc5253} demonstrated the great value of these type of
data at constraining properties such as the extinction in the MIR,
metallicity or stellar content (age, IMF, etc.).
 Here we present
observations of some locations in three young starburst
galaxies.
Two of these, the dwarf galaxy \iizw\ and the WR galaxy \henize\ are, together with NGC\,5253, well known
young starburst galaxies which possess embedded and compact knots
likely representing the earliest
evolutionary stages of super star clusters (SSCs)
or proto-globular clusters
\citep[e.g.][]{turner98,kobulnicky99,beck02,vacca02,johnson03}.
MIR images of these type of galaxies
\citep[e.g.][]{gorjian01,beck01} confirm the cluster hypothesis and indicate
that a large fraction ($\gtrsim 30-70$\%) of the total IR luminosity may originate from these compact knots. The third starburst galaxy we present here is NGC\,3256, a spiral galaxy which is the
brightest IR source in the nearby universe. MIR images \citep{boker97} reveal that NGC\,3256 has two distinct
nuclei aligned in the north-south direction and separated by
5\arcsec. The southern nucleus has no optical counterpart.
Hence, our observations include the two galactic nuclei of
NGC\,3256, the bright and compact infrared supernebula in \iizw\ and the two brightest infrared sources in the central
starburst of \henize\ \citep[named A and C following the nomenclature by][]{beck01}.

The core of this paper is structured as follows. Our observations are
described in Sect.~\ref{sect:observations}. Immediate results from our
TIMMI2 spectra are shown in Sect.~\ref{sect:results}. Here we present
for each object a general description, a discussion of the content of
its $N$-band spectrum and constraints on the extinction and
metallicity. Our main results are presented in
Sect.~\ref{sect:discussion}, where we discuss the importance of high
spatial resolution observations, make a comparison between the general properties of
these galaxies, examine the implications of the presence of PAH emission and
constrain the stellar content of the IR supernebula in \iizw. The main
conclusions of the paper are summarised in
Sect.~\ref{sect:conclusions}.

% XANDER {\bf The introduction misses how thhese galaxies ties in to the
%general scheme of things. Why should we study these galaxies? If you
%have seen one, have you seen them all? NO!}

\section{Observations and data reduction}
\label{sect:observations}

Our new infrared data on NGC\,3256, \iizw\ and \henize\ were obtained
as part of a program to observe young starburst galaxies with the
Thermal Infrared MultiMode Instrument (TIMMI2) on the ESO 3.6\,m
telescope (La Silla Observatory, Chile).

The $N$-band spectra of the nuclear regions of these three starburst
galaxies were obtained on 2003 March 20--26.  We used the 10~\micron\
low-resolution grism which ranges from 7.5 to 13.9~\micron\ and has a
spectral resolving power $\lambda/\Delta\lambda \sim 160$. The slit
used was 1\farcs2$\times$70\arcsec, with a pixel scale of 0\farcs45.
At the time of the observations, the slit could only be oriented in
the north-south direction. It was positioned across the bright
northern nucleus in the case of NGC\,3256 (including as well the southern
nucleus), across the bright infrared
compact core in the case of \iizw\ and across the two brightest MIR components in
the case of \henize, A and C (we refer to
Sects.~\ref{sect:ngc3256:general}, \ref{sect:iizw40:general}
and \ref{sect:he2-10:general} for a detailed description of the objects).
In order to correct for background emission from
the sky, the observations were performed using a standard
chopping/nodding technique along the slit in the north-south direction
(where the object is observed at two different positions on the slit)
with an amplitude of 20\arcsec\ in the case of NGC\,3256 and
10\arcsec\ in the cases of \iizw\ and \henize.  Adopted calibration
stars were HD\,90957 for NGC\,3256, HD\,37160 for \iizw\ and HD\,75691
and HD\,73603 for \henize.  They were observed right before and after
the targets and served as both telluric and flux standard stars.
 They were used as well for Point Spread Function (PSF) determination.
The synthetic calibrated spectra for these standard stars are given by
\citet{cohen99}.

The data processing included the removal of bad frames and the co-addition
of all chopping and nodding pairs. This left us with one single
image with one positive and two negative long-slit spectra.
These were combined with a simple shift-and-add
procedure which slightly increased the signal-to-noise.
For every individual observation, the spectra of the target and
standard star were
extracted using the optimal extraction procedure developed by
\cite{horne86}, ideal for unresolved or compact sources.
This procedure applies non-uniform pixel weights in the
extraction sum in order to reduce the statistical noise in the extracted
spectrum to a minimum while preserving its photometric accuracy.
The calibration of the spectroscopic data included 1) the  removal of the
telluric features, which was done by dividing by  the spectrum of
the standard star; 2) the removal of the spectral features of the
standard star; and 3) the absolute flux calibration. These last two
steps were achieved by multiplying by the synthetic spectrum of the
standard star.
 We propagate the uncertainty for each pixel, dominated by variations of
the sky transparency, along each step of the processing.

Wavelength calibration is straight forward since a
table with the pixel-to-wavelength correspondence is provided by the
TIMMI2 webpage\footnote{www.ls.eso.org/lasilla/sciops/timmi}.

We obtained two spectra of NGC\,3256, three of
\iizw, two of \henize\ A and three of the weaker \henize\ C. Each
spectrum was obtained after a total exposure time (on source) of $\sim
32$ min. The spectra of each source were combined to obtain one final
spectrum.

We compared the full-width-at-half-maximum (FWHM) measured along
the spatial direction for each object with the MIR seeing
(derived from the spatial FWHM of the calibration star)
in order to account for possible slit losses.

The FWHM along the spatial direction
measured for the southern nucleus (S) of NGC\,3256
is 1\farcs0$\pm$0\farcs2, which is of the order of the
MIR seeing (FWHM$\sim$0\farcs95--1\farcs11) and smaller than the slit
width (1\farcs2). Hence, slit losses towards NGC\,3256 S are
negligible. However, the northern nucleus (NGC\,3256 N) is slightly resolved with a FWHM
along the spatial direction of 1\farcs55$\pm$0\farcs05. Assuming that
the source is Gaussian and that the slit is perfectly centred on the
source peak, it implies that the slit might be registering about 68\%
of the total emission of the N nucleus.

Regarding the three long-slit spectra obtained of \iizw, the FWHM's measured
along the spatial direction
give values of 1\farcs67, 1\farcs03 and 1\farcs22, somewhat larger
than the average MIR seeing at the time of the observation (1\farcs11,
0\farcs91 and 0\farcs96, respectively). \iizw\ is then marginally
resolved but the slit is wide enough (1\farcs2) to ensure that
practically all the MIR flux emitted is registered. This is so since in the
last two cases the FWHM is similar to or smaller than the width of the
slit and the flux level of all three spectra agrees well within the
errors.

For \henize\ A, the FWHM's of the two spectra which were obtained
are 1\farcs83 and 1\farcs37,
larger than the respective MIR seeings (0\farcs77 and 0\farcs83) and the
slit width. Assuming that the source is Gaussian and that the slit is
perfectly centred on the source peak, it implies that the slit might
be registering about 60--75\% of the total flux emitted by component
A. In the case of \henize\ C, the three individual observations
obtained give FWHM's of 1\farcs22, 1\farcs13 and 1\farcs47, being the
respective MIR seeings 1\farcs0, 0\farcs88 and 1\farcs19. This source is
then only slightly resolved. As for \iizw, we are confident that the
slit is registering most of the MIR flux emitted by component C in
all three cases.

The spectra were extracted through apertures of about
$2\times{\rm FWHM}$ centred on the peak positions.

Line fluxes were measured by fitting a Gaussian and their quoted uncertainties (of the order of 5--20\%) include the statistical
error associated to each point at a given wavelength.
Upper limits are defined as the flux of a feature with a
peak flux three times the continuum rms noise and a width equal to the
instrumental resolution element.

\section{Results}
\label{sect:results}

Here we present our high spatial resolution spectroscopy data using
TIMMI2. First, a general description of every object is provided,
followed by a description of the spectra in terms of general shape and
line and molecular content. Our data is compared with observations
reported in the literature.

A variety of fine-structure lines and broad-band dust features fall
within the $N$-band spectroscopic range. The most relevant lines are
\ArIII\ at 9.0~\micron, \SIV\ at 10.5~\micron\ and \NeII\ at
12.8~\micron. These lines require hard radiation with energies between
$\sim$ 21 and 35~eV, and the most likely explanation for their
excitation mechanism is photoionisation by hot stars. In terms of
molecular features, the MIR spectra of many starbursts are
dominated by the well known emission features at 8.6, 11.2 and
12.7~\micron, commonly called the unidentified infrared (UIR) bands
and now generally attributed to vibrational emission of Polycyclic
Aromatic Hydrocarbons (PAHs) containing $\simeq 50$ carbon atoms
\citep[e.g.][]{allamandola89,puget89,peeters04:review}. A silicate
band centred around 9.7~\micron\ can also be present.

\subsection{NGC\,3256}
\label{sect:ngc3256}

\subsubsection{General description}
\label{sect:ngc3256:general}

The galaxy is a peculiar spiral galaxy located at a distance of
37~Mpc \citep{lipari00}. The angular scale at this distance is
1\arcsec = 179~pc. This galaxy consists of a main body of
about 60\arcsec, two extended tidal tails that can be traced as far as
8\arcmin\ and two faint external loops \citep{lipari00}. The central
($\sim$30\arcsec) region shows a knotty structure with a very unusual
triple asymmetrical spiral arm morphology. The double tidal tails are
characteristic of an interaction between two spiral galaxies of
comparable mass \citep[e.g.][]{deVaucouleurs61}.
%NGC\,3256 is the
%brightest IR source in the nearby universe (redshift $<
%3000$~km~s$^{-1}$), with a luminosity of $\sim
%3\times10^{11}\,L_{\sun}$ in the 8--1000~\micron\ range
%\citep{sargent89}. Molecular line studies \cite[e.g.][]{casoli92}
%indicate that the fraction of dense gas in NGC\,3256 is larger than in
%normal spiral galaxies, but smaller than in the ultra-luminous IR
%galaxy Arp\,220, which is one of the most embedded galaxies known so
%far.  HST H$\alpha$ images of NGC\,3256 by \cite{lipari04} show at
%least four giant galactic shells or bubbles, probably associated with
%explosions of supernovae and whose expansion likely began 7 Myr
%ago. These authors also find that 5\% of the total of young stellar
%cluster candidates might be massive clusters with masses in the range
%$\sim 10^6-10^7$\,$M_{\sun}$.

High spatial resolution near-IR
\citep[e.g.][]{moorwood94,kotilainen96}, MIR \citep{boker97} and
radio observations \citep[e.g.][]{norris95,neff03} reveal two distinct
nuclei aligned in the north-south direction and separated by
5\arcsec. The southern nucleus has no optical counterpart due to the
high extinction in this region. The presence of these two nuclei,
which at radio wavelengths have approximately similar size, brightness
and spectral index, suggests that the merger is not yet
completed. This is supported by the $K'$ surface brightness profile
over the central 15~kpc ($\sim$1\farcm4), which is clearly not that of
a galaxy that has already relaxed sufficiently to be classified as
elliptical \citep{moorwood94}. The radio spectral indices obtained for
both nuclei \citep[$\sim -0.9$;][]{norris95} are quite steep and indicate that the radio
emission is dominated by synchrotron radiation from cosmic-ray
electrons accelerated by supernovae and that the contribution to the
radio emission from \HII\ regions is negligible.

   \begin{figure}[!ht]
   \centering \includegraphics[width=8.8cm]{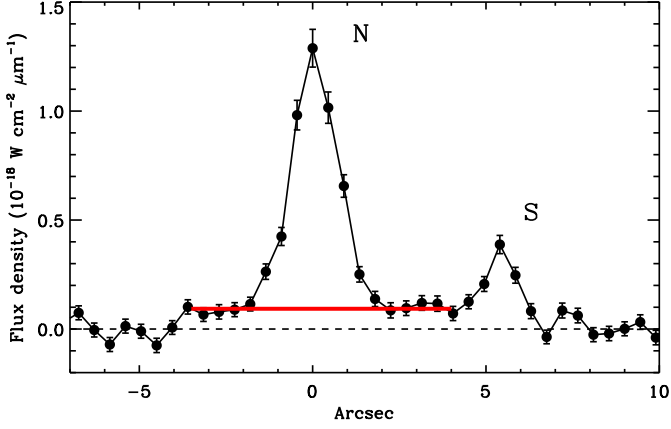}
   \caption{Spatial variation of the \NeII\ line peak across the slit
     from north (left) to south (right). Error bars indicate 1$\sigma$
     errors. A plateau which extends over
     $\sim 8$\arcsec is indicated by a solid line in light colour.  At
     the distance of NGC\,3256, 1\arcsec\ corresponds to $\sim 179$
     pc. The peak positions of the northern and southern nuclei are
     identified by the labels N and S respectively.}
         \label{fig:spatial}
   \end{figure}

The starburst nature of the northern nucleus of NGC\,3256 is indicated
by different authors : 1) The 10~\micron\ emission extends over a
region $\sim 4$~kpc ($\sim$22\arcsec) across with most of it
originating outside the central kpc ($\sim$5\farcs6) \citep{graham84}.
2) \cite{glass85} show that the $JHKL$ colours of the galaxy are
indicative of a young starburst.  3)  NGC\,3256 is located in the same
region as starburst galaxies in IR diagnostic diagrams which use PAH
emission features as a diagnostic tool for the ultimate physical
processes powering Galactic nuclei (AGN versus starburst), indicating
that at MIR wavelengths starburst activity is the dominant energy
source \citep{peeters04}.  4) NGC\,3256 also presents a prominent
3.3~\micron\ PAH feature \citep{moorwood86} with an equivalent width
typical of starburst galaxies.  5) \cite{rowan-robinson89} modelled
the IRAS 12, 25, 60 and 100~\micron\ fluxes and obtained that the
contribution of a hidden AGN is $<5$\%.  6) The specific search for
the [\ion{Si}{vi}] 1.96~\micron\ coronal line emission from a possibly
obscured AGN proved negative \citep{moorwood94}.  7) The central
3\arcsec\ of the nuclear region show strong recombination lines of
hydrogen and helium and a prominent CO band absorption at 2.3~\micron\
\citep{doyon94}. These features provide further evidence for starburst
activity since it implies the existence of a large number of OB stars
and a young population of red supergiants.  8) The SWS spectrum of
NGC\,3256 shows no signatures of high-excitation lines common in
galaxies powered by AGNs \citep{rigopoulou96}. This is confirmed by
the high spatial resolution spectrum of the northern nucleus obtained
by \cite{siebenmorgen04}.  The nature of the southern nucleus remains,
however, unclear.

   \begin{figure}[!ht]
   \centering \includegraphics[width=8.5cm]{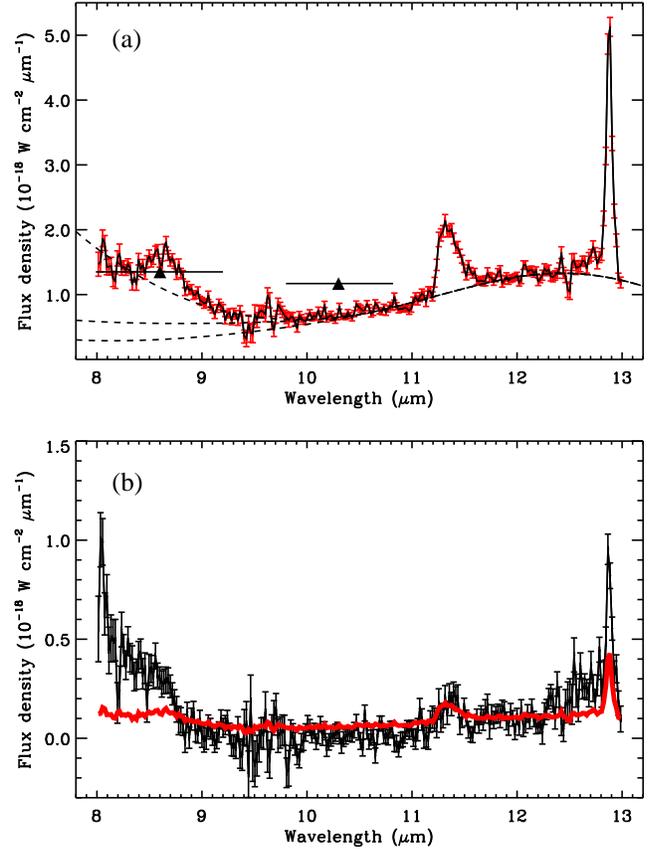}
   \caption{(a) $N$-band spectrum of the northern nucleus of
   NGC\,3256 with 1$\sigma$ errors.
   The spectrum is characterised by a strong dust
   continuum, the 8.6, 11.2 and possibly the 12.7~\micron\ PAH
   emission bands and the atomic fine-structure line of \NeII\ at 12.8
   \micron.  The dashed lines represent different local spline continua.
   The triangles indicate the TIMMI2 photometry in the
   8.6 and 10.4~\micron\ filters obtained by
   \cite{siebenmorgen04}. The FWHM's of these TIMMI2 filters are shown
   by horizontal bars.
   (b) $N$-band spectrum of the southern nucleus of NGC\,3256
   with 1$\sigma$ errors. The spectrum shows emission below $\sim
   8.8$~\micron, a rising dust continuum after $\sim 11$~\micron,
   the 11.2 \micron\ PAH emission band and the fine-structure line
   of \NeII\ at 12.8 \micron. For comparison, we show in light
   colour the spectrum of the northern nucleus scaled down by a
   factor of 12.}
         \label{fig:spec:north}
   \end{figure}

High spatial resolution (0\farcs6) Chandra observations \citep{lira02}
find several (14) discrete sources embedded in a complex diffuse
emission which contribute $\sim 20$\% of the total X-ray emission in
the 0.5--10~keV energy range. Two of these discrete sources are
coincident with the two nuclei, with the northern nucleus
corresponding to the brightest X-ray source. These authors find no
evidence for the presence of an AGN in the southern nucleus. However,
based on the ratio of 6~cm radio emission to 2--10 keV X-ray emission,
\cite{neff03} suggest the possibility that both nuclei might harbour
low-luminosity AGNs generated in or fuelled by the galaxy
merger. Nevertheless, based on HST STIS long-slit spectra of the
northern nucleus, \cite{lipari04} suggest that the shape of the
rotation curve and the emission-line profile can be explained by the
presence of young star clusters with outflow in the core, and that
most probably the associated compact X-ray emission and radio emission
are the result of a few recent supernovae remnants.

\subsubsection{N-band spectra}
\label{sect:ngc3256:spectrum}

The spectral observations towards the nuclear region of NGC\,3256
registered both the northern (N) and southern (S) nuclei previously
imaged at MIR wavelengths by \cite{boker97}. Fig.~\ref{fig:spatial}
shows the variation of the \NeII\ 12.8 \micron\ line peak flux across
the slit from north to south. The northern nucleus has a peak flux of
$(1.29\pm0.03) \times 10^{-18}$~W\,cm$^{-2}$\,\micron$^{-1}$. The
southern nucleus is much weaker, with a peak flux of $(0.39\pm0.03)
\times 10^{-18}$~W\,cm$^{-2}$\,\micron$^{-1}$. Faint \NeII\ emission
was also detected extending about 4\arcsec\ around the peak of
NGC\,3256~N. This faint emission appears as an emission plateau with
an average flux level of ($9.3\pm1.0) \times
10^{-20}$~W\,cm$^{-2}$\,\micron$^{-1}$. This plateau coincides with a
small spiral disc, at face-on position, which is a continuation of one
of the spiral arms of the galaxy and reaches the very northern nucleus
\citep{lipari00}.
% Further evidence of a nuclear disc is given by HST long-slit spectra
% of the northern nucleus, which show broadened H$\alpha$, H$\beta$,
% [\ion{N}{ii}] and [\ion{S}{ii}] emission lines with line widths of
% $\sim 450$~km~s$^{-1}$ \citep{neff03}. The emission lines indicate
% strong rotation around N with a peak velocity shear of $\sim
% 300$~km~s$^{-1}$ over a projected %distance of $\sim 40$~pc.

   \begin{figure}[!ht]
   \centering \includegraphics[width=8.8cm]{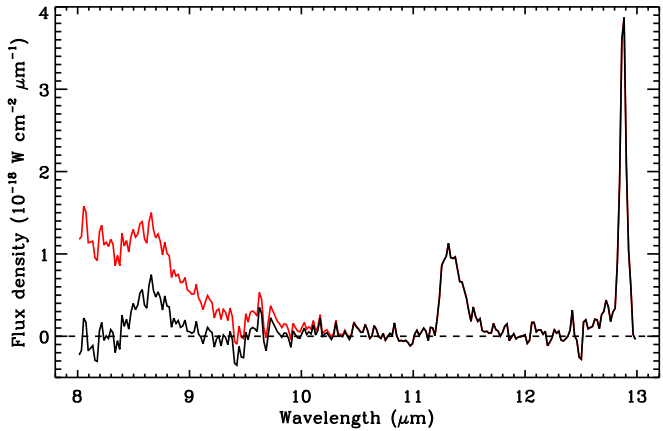}
   \caption{Continuum subtracted spectrum of NGC\,3256 N. The
   result of using the two extreme possibilities for the continuum
   plotted in Fig.~\ref{fig:spec:north} is shown by the dark and light
   lines. Note that the difference in both continua is the removal or
   not of the underlying PAH plateau and wing of the 7.7 \micron\,
   complex (see Fig.~\ref{fig:ir23133}).}
         \label{fig:spec:south}
   \end{figure}

   \begin{figure}[!ht]
   \centering \includegraphics[width=8.5cm]{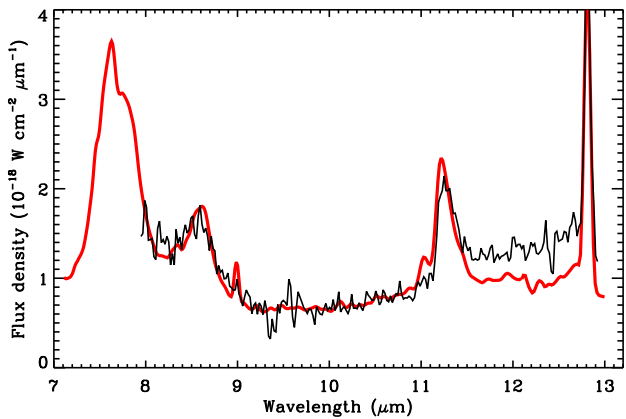}
   \caption{Comparison between the $N$-band spectrum of NGC\,3256~N
   (black line) and the ISO/SWS spectrum of the Galactic \HII\ region
   IRAS\,23133+6050 convolved to the resolution ($\lambda/\Delta
   \lambda \sim 160$) of TIMMI2 (light colour). The flux level of
   IRAS\,23133+6050 have been scaled down by a factor of 48. The
   wavelength array of NGC\,3256~N has been shifted by $-0.065$
   \micron\ in order to match the peak positions of the PAHs and
   \NeII\ line in the spectrum of IRAS\,23133+6050.}
         \label{fig:ir23133}
   \end{figure}

The fact that NGC\,3256~N is slightly resolved (see Sect.~\ref{sect:observations})
allows for a comparison
of the spatial variation of the \NeII\ line peak and other features
such as the 11.2~\micron\ PAH peak emission and the continuum at
12~\micron. No differences are evident, indicating that either the PDR
and dust shell form a thin layer around the N nucleus or we are simply
not resolving the emission.

Fig.~\ref{fig:spec:north}\,a shows the spectrum of NGC\,3256 N. Only
the valid range between 8 and 13~\micron\ is plotted. The spectrum is
characterised by a continuum due to warm dust, the 8.6, 11.2 and
possibly also the 12.7~\micron\ PAH emission bands, and the atomic
fine-structure line of \NeII\ at 12.8 \micron. \cite{siebenmorgen04}
have previously imaged the N nucleus at 8.6 and 10.4~\micron\ and
obtained a TIMMI2 spectrum of this nucleus with a 3\arcsec\ slit.
Their narrow-band filter photometry agrees well with the flux level of
our spectrum (cf. Fig.~\ref{fig:spec:north}\,a).  A comparison between
the spectrum obtained by \cite{siebenmorgen04} and ours in terms of
line and PAH band fluxes is done in Sects~\ref{sect:ngc3256:lines} and
\ref{sect:ngc3256:pahs}.

Emission towards the southern nucleus was only detected shortwards of
$\sim 8.8$~\micron\ and beyond 11~\micron.  The spectrum of NGC\,3256
S is shown in Fig.~\ref{fig:spec:north}b.  Like NGC3256 N, the
spectrum shows the 11.2~\micron\ PAH emission band, possibly also the
8.6 and 12.7~PAH bands, and the atomic fine-structure line of \NeII\
at 12.8 \micron. A comparison with the spectrum of the N nucleus is
also shown. Both spectra are similar, although with noticeable
differences in the emission shortwards of $\sim 8.8$~\micron\ and beyond
$\sim$ 12~\micron.
%{\bf I would say emission instead of continuum since
%this may well be a difference in PAH flux}

\cite{graham84} measured a 10~\micron\ flux of
$(5.1\pm0.4)\times10^{-18}$~W~cm$^{-2}$~\micron$^{-1}$ in a 15\arcsec\
aperture, about 8 times larger than the flux we obtain at this
wavelength towards the N nucleus, indicating that diffuse MIR
emission exists around the two nuclei
(cf. Sect.~\ref{sect:ngc3256:general}).

\subsubsection{Line fluxes}
\label{sect:ngc3256:lines}

Line fluxes towards NGC\,3256 N and S are listed in Table~\ref{table:ngc3256:fluxes}.
Only the line of \NeII\ at 12.8 \micron\ was detected. Upper limits to the
fluxes of the \ArIII\ 9.0 and \SIV\ 10.5 \micron\ lines are also
given. Considering the slit losses that
might be affecting the \NeII\ line flux emitted by NGC\,3256 N (about
32\%, see Sect.~\ref{sect:ngc3256:spectrum}), the total \NeII\ line flux would
be $\sim 31 \times 10^{-20}$~W~cm$^{-2}$.

Table~\ref{table:ngc3256:fluxes} compares the fluxes observed towards
the N and S nuclei with those observed by the large ISO/SWS beam and
other apertures/slits. The ISO/SWS aperture is approximately
14\arcsec$\times$20\arcsec\ up to 12~\micron\ and
14\arcsec$\times$27\arcsec\ between 12 and 19.6~\micron. The \ArIII\
line flux measured by ISO is approximately 4 times larger than our
upper limit, while the ISO \SIV\ flux is too small to be detected by
TIMMI2. Finally, the ISO \NeII\ line flux is about 3 times larger than
the combined flux emitted by the N and S nuclei. This indicates that
other objects besides the two nuclei must contribute strongly to the
\ArIII\ and \NeII\ line fluxes measured by ISO.

\cite{siebenmorgen04} only present the spectrum of the northern
nucleus and do not comment on the detection of the southern one,
probably due to the use of a wider slit (3\arcsec). Their published
spectrum most likely includes the contributions of both nuclei. They
estimate a \NeII\ line flux of $(28.5 \pm 1.0) \times
10^{-20}$~W~cm$^{-2}$, in good agreement with the value we obtain when
the fluxes of both nuclei are added together and slit losses are
considered.
Integrated over an aperture of 4\farcs2,
\cite{roche91} measured a \NeII\ line flux of
$38\times10^{-20}$~W~cm$^{-2}$, only slightly higher than the TIMMI2
line flux.

\subsubsection{PAH bands}
\label{sect:ngc3256:pahs}

Both nuclei exhibit PAH emission features (see
Fig.~\ref{fig:spec:north}). The northern nucleus clearly shows the 8.6
and 11.2~\micron\ PAH bands, possibly the 12.7 \micron\, PAH band, the
red wing of the 7.7~\micron\ PAH complex and that of the broad
emission plateau underlying both the 7.7~\micron\ complex and the 8.6
\micron\ band, clearly seen shortwards of $\sim$9~\micron\ \citep[see
Fig.~\ref{fig:ir23133} for a comparison with the ISO/SWS spectrum
of the Galactic \HII\ region
IRAS\,23133$+$6050 and e.g. the review by][]{peeters04:review}.  The southern
nucleus shows the 11.2~\micron\ band and possibly also the 8.6 and
12.7~\micron\ PAH bands.

\begin{table}[!ht]
\caption{Line fluxes and $3\sigma$ upper limits measured towards the two nuclei in NGC\,3256. Also included are line fluxes measured within other aperture/slits.}
  \label{table:ngc3256:fluxes}
  \begin{center}
    \leavevmode
    \begin{tabular}[h]{lccccc}
      \hline \hline \\[-7pt]
    \multicolumn{1}{c}{Line} &
    \multicolumn{1}{c}{$\lambda$} &
    \multicolumn{1}{c}{North} &
    \multicolumn{1}{c}{South} &
    \multicolumn{1}{c}{Aperture} &
    \multicolumn{1}{c}{Ref.}\\
    \multicolumn{1}{c}{} &
    \multicolumn{1}{c}{(\micron)} &
    \multicolumn{2}{c}{($10^{-20}$~W~cm$^{-2}$)} &
    \multicolumn{1}{c}{(\arcsec)} &
    \multicolumn{1}{c}{}\\[5pt] \hline \\[-7pt]

\ArIII       & 9.0  & $<1.2$       & $<1.2$            & 1.2 & 1\\
\SIV         & 10.5 & $<1.1$       & $<1.1$            & 1.2 & 1\\
\NeII        & 12.8 & $21 \pm 1^\star$ & $3.3 \pm 0.6$ & 1.2 & 1\\[3pt]

\SIV         & 10.5 & \multicolumn{1}{c}{$<1.6^\lozenge$} & & 3 & 2\\
\NeII        & 12.8 & \multicolumn{1}{c}{$28.5 \pm 1.0^\lozenge$} & & 3  & 2\\[3pt]

\NeII        & 12.8 & $38^\lozenge$ & & 4.2 & 3 \\[3pt]

\ArIII       & 9.0  & \multicolumn{2}{c}{$4.7 \pm 0.4^\triangle$} & $14\times20$ & 4\\
\SIV         & 10.5 & \multicolumn{2}{c}{$0.9 \pm 0.1^\triangle$} & $14\times20$ & 4\\
\NeII        & 12.8 & \multicolumn{2}{c}{$89 \pm 14^\triangle$}   & $14\times27$ & 4\\[3pt]

\hline

    \end{tabular}
  \end{center}
$^\star$ When correcting for slit losses, we obtain a total line flux of about
$31 \times 10^{-20}$~W\,cm$^{-2}$ (cf. Sect.~\ref{sect:ngc3256:spectrum}).\\
$^\lozenge$ This observation is probably centred on the N nucleus.\\
$^\triangle$ The ISO/SWS aperture includes both N and S nuclei and any other possible mid-IR emitters. Errors are calculated from the
typical ISO/SWS $1\sigma$ absolute flux accuracy \citep{peeters:catalogue}.\\
REFERENCES:
(1) This work;
(2) \cite{siebenmorgen04};
(3) \cite{roche91};
(4) \cite{verma03}.

\end{table}

\begin{table}[!ht]
\caption{Continuum subtracted 8.6, 11.2 and 12.7~\micron\ PAH band
  fluxes (in units of $10^{-20}$ W~cm$^{-2}$) measured towards the
  northern and southern nuclei. These are compared with other observations
  at different apertures. }
  \label{table:ngc3256:pahs}
  \begin{center}
    \leavevmode
    \begin{tabular}[h]{ccccc}
      \hline \hline \\[-7pt]
    \multicolumn{1}{c}{PAH band} &
    \multicolumn{1}{c}{North} &
    \multicolumn{1}{c}{South} &
    \multicolumn{1}{c}{Aperture} &
    \multicolumn{1}{c}{Ref.}\\
    \multicolumn{1}{c}{} &
    \multicolumn{2}{c}{($10^{-20}$~W~cm$^{-2}$)} &
    \multicolumn{1}{c}{(\arcsec)} &
    \multicolumn{1}{c}{}\\[5pt] \hline \\[-7pt]

8.6  & $18\pm4$   & --                  & 1.2 & 1\\
11.2 & $25\pm4$   & $5\pm2$             & 1.2 & 1\\
12.7 & $<13$      & $<7$                & 1.2 & 1\\[5pt]

11.2  & 30         & --                  & 3   & 2\\
11.2  & 32$^\star$ & --                  & 4.2 & 3\\
11.2  & \multicolumn{2}{c}{107$^\sharp$} & 24  & 2\\[5pt]

\hline

    \end{tabular}
  \end{center}
$^\star$ Although it is not indicated by the authors, we assume that these observations are centred on the N nucleus.\\
$^\sharp$ The ISOPHOT aperture includes both N and S nuclei and any other possible mid-IR emitters.\\
REFERENCES:
(1) This work;
(2) \cite{siebenmorgen04};
(3) \cite{roche91}.
\end{table}

%%  PAH emission features are clearly seen at 11.2~\micron\ in both nuclei
%%  (see Fig.~\ref{fig:spec:south}).  The northern nucleus also shows a
%%  prominent 8.6~\micron\ band, situated on top of the broad emission
%%  plateau underlying both the 7.7~\micron\ complex and the 8.6 \micron\
%%  band,
%%  %e.g. IRAS\,23133$+$6050 in Fig.~\ref{fig:ir23133} or
%%  \citep[cf.][]{peeters04:review}, and possibly also the 12.7 \micron\,
%%  band (Fig.~\ref{fig:spec:north}).  This band might also be present in
%%  the southern nucleus (Fig.~\ref{fig:spec:south}).

The PAH fluxes are determined by subtracting a local spline continuum
(e.g. Fig.~\ref{fig:spec:north}) and are given in Table
\ref{table:ngc3256:pahs}. The presence of the 12.7 \micron\ band is
highly dependent on the choice of the continuum which, in turn, is
influenced by the flux level at the longest wavelengths (around
13~\micron).
Therefore, its detection is merely tentative and thus we simply give
an upper limit.

The 11.2~PAH flux we obtain towards the N nucleus agrees well with
those obtained through slightly larger apertures (3\arcsec-- 4\arcsec,
see Table~\ref{table:ngc3256:pahs}).  However, when compared with the
11.2~PAH flux measured by the large ISOPHOT aperture, the combined
flux of both the N and S nuclei accounts only for 30\% of the
total. This is roughly the same contribution of the \NeII\ flux
emitted by both nuclei to the total ISO/SWS \NeII\ flux. This suggests
that this non-nuclear emission comes from \HII\ regions and their
associated PDRs.

The comparison of the PAH emission bands in both nuclei (see
Fig.~\ref{fig:spec:north}) reveals large differences in both the
relative strength of the 11.2/12.7 PAH band ratio (assuming for the
moment a positive detection of the 12.7 \micron\, band in both nuclei)
and the plateau/11.2 ratio. In both cases, the southern nucleus has a
stronger 12.7 PAH band and a stronger plateau both relative to the
11.2 \micron\ PAH band compared to that of the northern nucleus.
Possible origins for this are the following. (i) A different PAH
charge balance in both nuclei. Indeed, a larger fraction of ionic PAHs
give rise to stronger emission in the 5-10 \micron\ region
\citep[e.g.][]{Peeters02} and hence might explain the difference in
plateau/11.2 ratio. For the 12.7 \micron\ PAH, it is unclear at this
moment whether it is due to neutral or cationic PAHs
\citep{hony01}. (ii) The 12.7/11.2 ratio is a tracer of the PAH
molecular edge structure \citep{hony01}. In this case, the S nucleus
should have more irregular PAHs \citep{hony01} likely due to a
stronger radiation field and lower density. Molecular edge structure
has little influence on the CC stretching and CH in-plane-bending
modes emitting in the 6-9 \micron\ region. (iii) The underlying
plateau is believed to arise from larger PAH-related molecules or
complexes \citep{allamandola89} indicating that both nuclei might have
a different PAH size distribution. As with the other two possibilities,
it cannot explain at once the relative intensities of all PAH ratios.

\subsubsection{Extinction}
\label{sect:ngc3256:extinction}

The extinction suffered by the line emission from NGC\,3256~N has been
estimated by \cite{doyon94} using two different methods: 1) the ratio
of Pa$\beta$ to Br$\gamma$, which gives an absolute extinction at
2.2~\micron\ of $A_{\rm K}=0.55\pm0.09$~mag; and 2) the ratio of
[\ion{Fe}{ii}] 1.257 to [\ion{Fe}{ii}] 1.644~\micron, which implies
$A_{\rm K}=0.7\pm0.3$~mag. \cite{kotilainen96} proposed a third
method, which assumes the correlation between 6~cm radio and
[\ion{Fe}{ii}] luminosities established by \cite{forbes93}. They
obtain an extinction of $A_{\rm K}=0.51$ for the N nucleus and of
$A_{\rm K}=1.1$ for the S nucleus. For the N nucleus, the average of
the three methods gives $A_{\rm K}=0.6$. These values of the
extinction in the near-IR, i.e. $A_{\rm K}$(N)$=0.6$ and $A_{\rm K}$(S)$=1.1$,
translates into $A_{\rm V}\sim 5.6$ and
10.2~mag for the N and S nuclei, respectively, assuming $R_{\rm
V}=3.1$ \citep{mathis90}. Based on the $H-K$ colours, \cite{lipari00}
find $A_{\rm V}$(N)$\sim 5.5$ and $A_{\rm V}$(S)$\sim 16$~mag.

The extrapolation of these values of the extinction in the optical and
near-IR to the MIR regime is not direct unless the exact shape and
depth of the silicate absorption feature at 9.7~\micron\ is
known.
%{\bf XANDER: what about Jean Chiar's extinction law?}
This could be done by fitting the
dust continuum \cite[cf.][]{martin:ngc5253}. However, the presence of
the strong PAH bands hampers this task considerably. By simply
assuming that the extinction law in the MIR can be described by the
commonly used ``astronomical silicate''
\citep[cf.][]{draine84,draine85}, we obtain that $A_{12.8}=0.33A_{\rm
sil}$ \citep{martin:atca:gal}, where $A_{12.8}$ is the extinction at
the \NeII\ line central wavelength and $A_{\rm sil}$ is the extinction
at the peak of the silicate absorption. Assuming a ratio $A_{\rm
V}/A_{\rm sil}\sim 18.5$, found by \cite{roche84} for the local
diffuse interstellar medium, we obtain $A_{\rm sil}$(N)$\sim 0.4$ and
$A_{\rm sil}$(S)$\sim 0.6$~mag when the values above mentioned for the
V-extinction towards the N and S nuclei (5.6 and 10.2) are
used. Hence, we obtain $A_{12.8}$(N)$\sim 0.1$ and $A_{12.8}$(S)$\sim
0.2$~mag. These values are to be considered as only rough estimates of
the extinction affecting the \NeII\ line. When applying these
extinction values to the \NeII\ line fluxes measured for both nuclei,
we obtain $\sim 23\times10^{-20}$ and $\sim
4\times10^{-20}$~W~cm$^{-2}$, respectively. These fluxes are still
within the 2$\sigma$ error bars of the observed fluxes. Even in the
case of a larger extinction towards the S nucleus (16~ mag, see
above), the \NeII\ extinction corrected flux will be $\sim
4.3\times10^{-20}$, only 30\% larger than the observed flux.

%In Fig.~\ref{fig:ir23133} we compare the spectrum of NGC\,3256~N with the ISO/SWS spectrum %the Galactic \HII\ region IRAS\,23133$+$6050 \citep[cf.][]{peeters:catalogue}. The similitude %between both spectra is striking. IRAS\,23133$+$6050 is a spherical \HII\ region which %suffers little or no extinction in the near-IR ($A_{\rm K}$ is close to 0) as determined by
%the ratio of Br$\beta$ to Br$\alpha$ \citep[cf.][]{martin:paperii}. It does not show %indications of a silicate absorption neither. On basis of the spectral similitude, we can %argue that the N nucleus must indeed be little affected by extinction in the mid-IR.

%Since the exact shape of the extinction curve in the mid-IR for both
%nuclei cannot be directly estimated from our TIMMI2 spectrum and it
%seems that its effect on the line fluxes is not large, we will not
%attempt to apply any extinction correction to the fluxes listed in
%Tables~\ref{table:ngc3256:fluxes} and \ref{table:ngc3256:pahs}.

\subsubsection{Ionic abundances}
\label{sect:ab:ngc3256}

Ionic abundances can be determined from the measured strengths of the
fine-structure and \HI\ recombination lines.
The ionic abundance of a certain ion $X^{\rm{+i}}$ with respect to hydrogen
($X^{\rm +i}/{\rm H^+}$) can be determined using the following
expression \citep[e.g.][]{rubin88}:

\begin{equation}
{X^{\rm +i} \over {\rm H^+}} = { {F_{X^{\rm +i}}/F_{{\rm HI}}} \over
{\epsilon_{X^{\rm +i}}/\epsilon_{{\rm HI}}} }~,
\label{eq:ab}
\end{equation}

\noindent
where $F_{X^{\rm +i}}$ and $F_{{\rm HI}}$ are the extinction-corrected
fluxes of any line produced by $X^{\rm{+i}}$ and \HI, and
$\epsilon_{X^{\rm +i}}$ and $\epsilon_{{\rm HI}}$ are their respective
emission coefficients. This expression assumes that 1) the nebula is
homogeneous with constant electron temperature and density; 2) all the
line photons emitted in the nebula escape without absorption and
therefore without causing further upward transitions; and 3) the
volume occupied by $X^{\rm{+i}}$ and H$^+$ is the same.  The
fine-structure line emission coefficients depend on the electron
temperature (\tel), density (\den) and relevant atomic parameters
\citep[cf.][]{martin:paperii}. The \HI\ emission coefficients can be
derived by using the program \textsc{intrat} by \cite{storey95}.

In the case of NGC\,3256, we can determine the abundance of Ne$^+$
with respect to H$^+$.  Adopting \tel$=6000$~K
\citep{aguero91,storchi95,lipari00} and \den$=10^3$~\cm\
\citep{lipari00} for both the N and S nuclei, we obtain $\epsilon_{\rm
[NeII]\,12.8}=9.95\times10^{-22}$ and $\epsilon_{\rm
Br\gamma}=5.98\times10^{-27}$~erg~\cm~s$^{-1}$.  \cite{kotilainen96}
measured the Br$\gamma$ flux within 3\arcsec\ apertures centred
on both nuclei. They obtain de-reddened
fluxes of $0.52\times10^{-20}$ and $0.14\times10^{-20}$~W~cm$^{-2}$
respectively for the N and S nuclei.
We note that the Br$\gamma$ line
fluxes measured by \cite{doyon94}, integrated over a square aperture
of 3\farcs5$\times$3\farcs5, are roughly twice those obtained by
\cite{kotilainen96}. However, we consider here the fluxes by
\citeauthor{kotilainen96} since their Br$\gamma$ flux towards the N
nucleus predicts well the extinction corrected H$\beta$ flux measured
by \cite{lipari00}. We will consider a standard 20\% uncertainty for these line fluxes.

Using the \NeII\ line fluxes listed in
Table~\ref{table:ngc3256:fluxes}, we obtain Ne$^+$/H$^+$(N)$\gtrsim
[2.4(2.6)\pm0.5]\times10^{-4}$ and Ne$^+$/H$^+$(S)$\gtrsim[1.4(1.7)\pm0.5]\times10^{-4}$,
 where the values in brackets are obtained by using the extinction-corrected \NeII\ line
fluxes given in Sect.~\ref{sect:ngc3256:extinction}.
The use of the Br$\gamma$ flux
measured by \cite{doyon94} will give Ne$^+$ abundances twice as large. There also exists the fact that the \NeII\ and Br$\gamma$ lines are not measured on the same aperture.
In principle, one could be over/underestimating the Br$\gamma$ flux associated to the MIR sources. This
uncertainty is, however, difficult to quantify.

Formally, the above computed Ne$^+$ abundances are lower limits to
the Ne abundance: Ne/H(N)$\gtrsim [2.0(2.2)\pm 0.4]\times$[Ne/H]$_{\sun}$ and
Ne/H(S)$\gtrsim[1.2(1.7)\pm 0.4]\times$[Ne/H]$_{\sun}$.
These limits agree with previous optical determinations
\citep{lipari00,aguero91,moran99}
which, using different calibrations, find oxygen abundances of ~1.3 solar
or even 2.6--3 times solar
\citep[assuming $12+\log$(O/H)$_{\sun}=8.69$,][]{allende01}.
However, several uncertainties affect the oxygen abundances derived
from emission lines using the  "strong line method" especially
at high metallicity \citep[e.g.][]{pilyugin01,stasinska05}.
%(e.g. Pilyugin 2001, Stasinska 2005, A&A 434, 507).
Furthermore, if the latest upward revision of the solar Ne abundance
by $\sim$0.4--0.5~dex suggested from helioseismology \citep{antia05,bahcall05}
and from solar type stars \citep{drake05} is confirmed,
our Ne abundance estimate could well be compatible with solar or even
slightly sub-solar.

\subsection{II\,Zw\,40}
\label{sect:iizw40}

\subsubsection{General description}
\label{sect:iizw40:general}

\iizw\ is at a distance of 9.2~Mpc \citep{vacca92}. This distance
accounts for Galactic rotation but no Virgocentric flow, and assumes
$H_0=75$~km~s$^{-1}$~Mpc$^{-1}$. At this distance, 1\arcsec\
corresponds to 44.6~pc.

This compact galaxy consists of a bright core only
3\arcsec$\times$5\arcsec\ in size with two faint and diffuse tails
extending about 30\arcsec\ to the south-east
\citep{sargent70}. \cite{baldwin82} carried out an emission-line
optical study of the morphology and kinematics of \iizw. They
considered that the two tails are reminiscent of interacting galaxies
and suggested a possible collision between two small systems as the
source of the present starburst.

\iizw\ is one of the prototype of the class of dwarf galaxies known as
``detached extra-galactic \HII\ regions'' \citep{sargent70}. These
systems are characterised by emission-line spectra
\citep[e.g.][]{walsh93} similar to those of individual giant \HII\
regions superimposed on a continuum which appears to be mainly due to
O stars.
%Indeed, near-IR line and continuum measurements indicate that
%nebular recombination emission and photospheric radiation from young
%blue stars produce most of the near-IR continuum emission in the
%central 6\arcsec\ \citep{joy88}. Evolved stars, which dominate the
%near-IR emission from normal galaxies, contribute no more than 25\% of
%the total 2.2~\micron\ flux in the central region of \iizw, with a
%stellar mass ratio between O, B and A type stars and evolved stars
%$M_{\rm OBA}/M_{\rm evolved} \sim 0.1$, exceptionally large compared
%to the nominal $M_{\rm OBA}/M_{\rm evolved} \sim 10^{-4}$ found in
%normal galaxies \citep[e.g.][]{thronson88}.
The presence of about 200
WR stars is inferred from the flux of the \HeII\ 4684~\AA\ line
\citep[cf.][]{vacca92}.

   \begin{figure}[!ht]
   \centering \includegraphics[width=8.5cm]{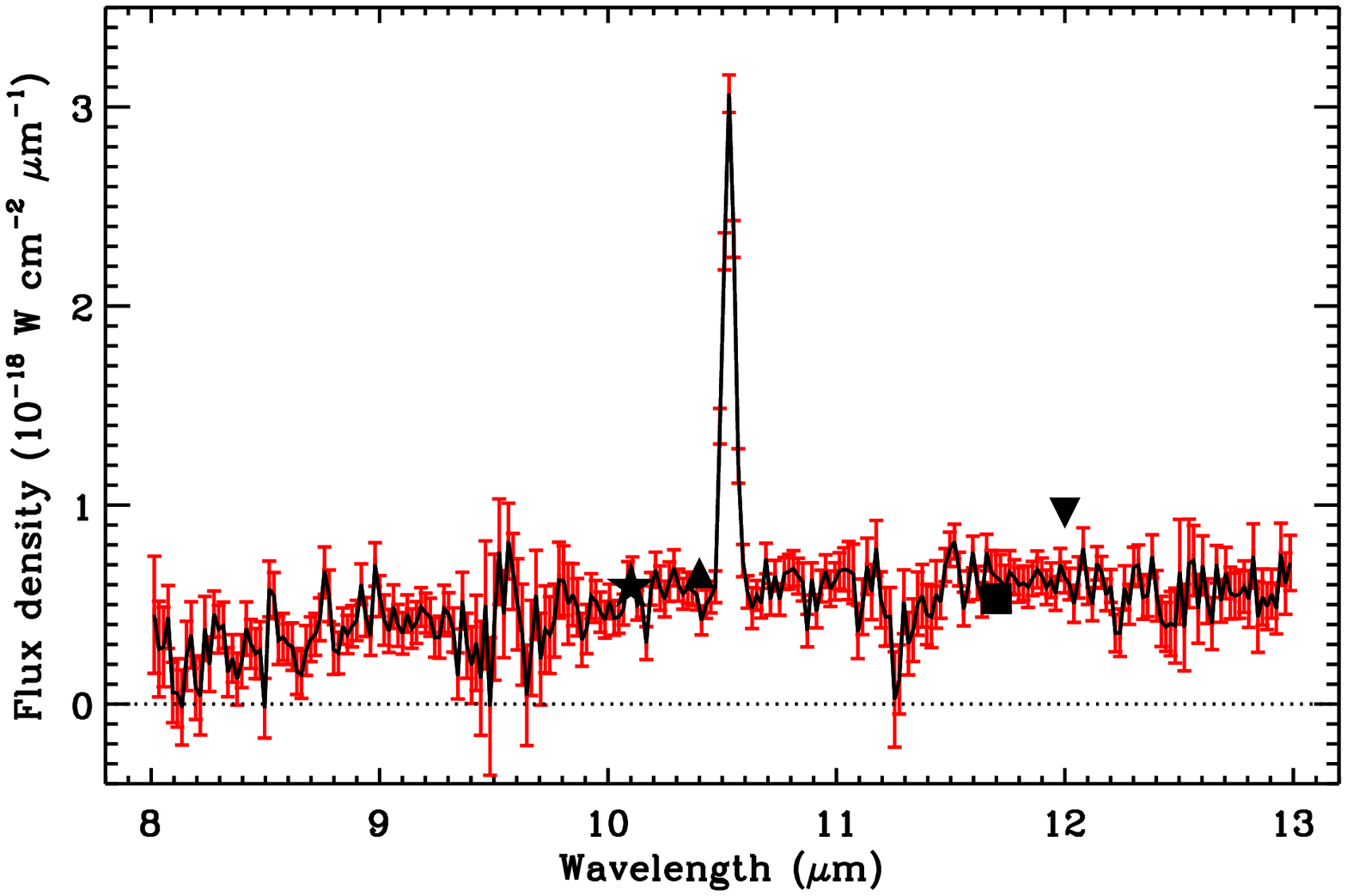}
   \caption{$N$-band spectrum of the infrared supernebula in
   II\,Zw\,40 with 1$\sigma$ errors.  The spectrum is characterised by a rather flat
   continuum and a strong \SIV\ 10.5~\micron\ line. For comparison, we
   plot the 10.1~\micron\ flux measured by \cite{wynn86} in a 7\farcs7
   aperture (star), the 10.4~\micron\ flux measured by \cite{rieke72}
   in a 6\arcsec\ aperture (triangle), and the IRAS 12~\micron\ flux
   \citep[upside down triangle;][]{vader93}. All these are broad band
   (filter FWHM $\sim 5$~\micron) photometric fluxes.  The narrow band
   filter 11.7~\micron\ flux measured by \cite{beck02}, with a FWHM of
   $\sim 1$~\micron, is indicated by a square.}
         \label{fig:iizw40}
   \end{figure}

\begin{table}[!ht]
\caption{Line fluxes and 3$\sigma$ upper limits measured towards \iizw. These are compared
	with measurements made with ISO/SWS.}
  \label{table:fluxes:iizw40}
  \begin{center}
    \leavevmode
    \begin{tabular}[h]{lcccc}
      \hline \hline \\[-7pt]
    \multicolumn{1}{c}{Line} &
    \multicolumn{1}{c}{$\lambda$} &
    \multicolumn{1}{c}{Line Flux} &
    \multicolumn{1}{c}{Aperture} &
    \multicolumn{1}{c}{Ref.} \\
    \multicolumn{1}{c}{} &
    \multicolumn{1}{c}{(\micron)} &
    \multicolumn{1}{c}{($10^{-20}$ W cm$^{-2}$)} &
    \multicolumn{1}{c}{(\arcsec)} &
    \multicolumn{1}{c}{}
    \\[5pt] \hline \\[-7pt]

\ArIII &  9.0 & $< 2.6$        & 1.2 & 1\\
\SIV   & 10.5 & $15.5 \pm 0.8$ & 1.2 & 1\\
\NeII  & 12.8 & $< 2.2$        & 1.2 & 1\\[5pt]

\ArIII &  9.0 & $2.0 \pm 0.2$ & $14\times20$ & 2\\
\SIV   & 10.5 & $20 \pm 2$    & $14\times20$ & 2\\
\NeII  & 12.8 & $1.7 \pm 0.3$ & $14\times27$ & 2\\[5pt]

\hline

    \end{tabular}
  \end{center}
REFERECES:
(1) This work;
(2) \cite{verma03}.
\end{table}

The radio continuum emission of \iizw\ is relatively compact and
coincides with the optical bright core
\citep[e.g.][]{wynn86,sramek86,klein91,deeg93,beck02}. The spectral
index of the overall radio emission has been discussed at length. It
seems that the total cm-wave emission of \iizw\ is mostly thermal,
with at most 30\% of the total radio emission attributed to non-thermal
processes \citep[e.g.][]{sramek86,joy88}. Within the central 6\arcsec,
the observed spectral index equals or exceeds the value for optically
thin free-free emission ($\alpha \gtrsim -0.1$, with $S_{\nu} \propto
\nu^{\alpha}$), indicating no evidence for non-thermal emission from
supernovae events and remnants. Further evidence of this lack of
supernovae comes from the faintness of the [\ion{Fe}{ii}] line emission
at 1.644~\micron\ \citep[e.g.][]{vanzi96}, which is more likely to be
produced in shocks than in photoionised gas. High-spatial resolution
($\sim$ 0\farcs1) imaging at 2~cm reveals a complex of smaller nebulae
within the bright radio core \citep{beck02}.

\iizw\ was first detected at 10~\micron\ by \cite{rieke72}. Later
observations by \cite{wynn86} reveal that most of the 10~\micron\
emission arises from a region less than 4\arcsec\ in diameter, more
compact than the near-IR and optical emission. Recently, Keck MIR
imaging has shown a bright source with a diameter (FWHM) of about
0\farcs5 coincident with the radio compact emission \citep{beck02}.

\subsubsection{N-band spectrum}
\label{sect:iizw40:spectrum}

Fig.~\ref{fig:iizw40} shows the spectrum of the infrared nebula in
\iizw. Only the valid range between 8 and 13~\micron\ is
plotted. The spectrum of \iizw\ is characterised by a rather flat
continuum and a strong \SIV\ line at 10.4~\micron.

\iizw\ has been previously observed at MIR wavelengths using a wide
range of apertures: at 10.1~\micron\ in a 7\farcs7 aperture
\citep{wynn86}; at 10.4~\micron\ in a 6\arcsec\ aperture; at
11.7~\micron, with a spatial resolution of 0\farcs3--0\farcs5; and at
12~\micron\ with IRAS \citep{vader93}. These various photometric
fluxes are indicated in Fig.~\ref{fig:iizw40} and show a perfect
agreement with the spectrum we have obtained, even in the case of the
IRAS 12~\micron\ flux measured with a beam that includes the entirely
galaxy. This suggests that the MIR flux of the galaxy is confined
to the compact (0\farcs5) infrared source we have observed with
TIMMI2.

\subsubsection{Line fluxes}
\label{sect:iizw40:lines}

%The $N$-band spectral range includes the fine structure lines of \ArIII\ 9.0, \SIV\ 10.4 and %\NeII\ 12.8~\micron.
Towards \iizw, only the \SIV\ line is detected. Its line flux and the
upper limits of the non-detected lines are measured as described above
(cf. Sect.~\ref{sect:ngc3256:lines}) and listed in
Table~\ref{table:fluxes:iizw40}

Table~\ref{table:fluxes:iizw40} compares the fluxes observed by TIMMI2
towards the infrared nebula in \iizw\ with those observed by the large
ISO/SWS aperture. The upper limits to the \ArIII\ and \NeII\ line
fluxes are consistent with the ISO values. Regarding \SIV, the
ISO flux is only 1.3 times larger than the TIMMI2 value. Hence, it
seems valid to consider (as it has been stated in the previous
section) that practically all of the mid-infrared line emission
emitted by the galaxy comes from this compact nebula.

\subsubsection{Extinction}
\label{sect:iizw40:extinction}

The analysis of the extinction towards \iizw\ is complicated by the
low galactic latitude of the galaxy, which implies some foreground
extinction. Values of the extinction based on a foreground screen
model and derived from the H$\alpha$/H$\beta$, H$\alpha$/Br$\gamma$ or
H$\beta$/Br$\gamma$ line ratios are in the range $A_{\rm V}=
2.1-3.0$~mag
\citep[e.g.][]{sargent70,wynn86,vanzi96,joy88,davies98}. \cite{walsh93}
present a map of the total (Galactic + extra-galactic) $A_{\rm V}$
derived from the H$\alpha$/H$\beta$ ratio. The map contains values of
$A_{\rm V}$ as large as 4.5~mag, found preferentially in the northern
half of the bright optical core of \iizw\ with a general decrease to
the south-east. However, non optical studies obtain much larger visual
extinctions.
Near-IR observations of \iizw\ give Br$\gamma$ fluxes of $(3.40-4.55)\times10^{-21}$ W\,cm$^{-2}$ measured in apertures between 3\arcsec and 9\arcsec
\citep[][]{joy88,ho90,vanzi96,davies98,coziol01}. Using this range in the observed Br$\gamma$ flux, \cite{beck02} estimated an extinction of $A_{\rm V} \sim 8-10$ based on their 2 cm free-free flux from a region $\sim 3$\arcsec.
%They explain this difference
%as probably being caused by the fact that much of the extinction is
%high and internal to the nebula.
A similar value for the visual
extinction ($A_{\rm V} \sim 10$~mag) is found by \cite{verma03} based
on ISO \HI\ recombination lines. Other authors also argue (though from
modelling of the SED) for a very high extinction ($A_{\rm V} \sim 20-30$) for
this object \citep{hunt05}. Such a high extinction should lead to a significant silicate absorption band, which is not seen in the TIMMI2 spectrum (Fig.~\ref{fig:iizw40}). Hence, unless the dust is very deficient in silicate, we do not favor such a high $A_{\rm V}$.
\cite{jaffe78} propose a mixture of
foreground and internal extinction and reproduce the Balmer decrement
with foreground and internal extinctions of 0.8 and 8 mag respectively.

The extrapolation of the extinction in the optical and near-IR to the
MIR regime is not direct. We can, however, estimate upper limits to
the extinction that affects the MIR lines observed in \iizw\ by
adopting $A_{\rm V}=10$~mag ($A_{\rm K}=1.1$, \citeauthor{rieke85} \citeyear{rieke85})
and the method described in
Sect.~\ref{sect:ngc3256:extinction}, i.e. using an extinction law defined
by the "astronomical silicate" with $A_{\rm V}/A_{\rm sil}=18.5$.
We obtain
$A_{9.0} =0.86\times A_{\rm sil}=0.46$,
$A_{10.5}=0.80\times A_{\rm sil}=0.43$,
$A_{12.8}=0.33\times A_{\rm sil}=0.18$,
$A_{15.5}=0.27\times A_{\rm sil}=0.15$,
$A_{18.7}=0.36\times A_{\rm sil}=0.19$ and
$A_{25.9}=0.19\times A_{\rm sil}=0.10$~mag,
where these are, respectively, the extinction at
the central wavelengths of the \ArIII\ 9.0, \SIV\ 10.5, \NeII\ 12.8,
\NeIII\ 15.5, \SIII\ 18.7 and \OIV\ 25.9~\micron\ lines
\citep[cf.][]{martin:atca:gal}. The extinction affecting the \ArII\
7.0~\micron\ is obtained by simply extrapolating the value in the
$K$-band considering a power law $A_{\lambda} \propto
\lambda^{-1.7}$. We obtain $A_{7.0}=0.14 \times A_{\rm K} = 0.15$~mag.

These values of the extinction have to be considered as simply rough
estimates and will be later considered in Sect.~\ref{sect:cloudy},
where the stellar content of \iizw\ will be discussed. Moreover, we might be
overestimating the extinction for the lines around 10 and 18\,\micron\ since, as we mention above, the TIMMI2 spectrum does not show evidence for silicate absorption.

\subsubsection{Ionic abundances}
\label{sect:ab:iizw40}

Following Sect.~\ref{sect:ab:ngc3256}, we can determine the ionic
abundance of S$^{3+}$ with respect to H$^+$.
\cite{beck02} predict an unextincted
Br$\gamma$ flux of $1.2\times10^{-20}$~W~cm$^{-2}$ based on radio continuum observations (see previous section). We will use this value hereafter and consider a standard 20\% uncertainty.

Adopting \tel$=12500$~K \citep{vacca92,walsh93,perez03} and \den$=1700$~\cm\
(see Table~\ref{table:prop}), we obtain $\epsilon_{\rm
[SIV]\,10.5}=5.07\times10^{-20}$ and $\epsilon_{\rm
Br\gamma}=2.66\times10^{-27}$~erg~\cm~s$^{-1}$. The resulting
S$^{3+}$/H$^+$ is $[6.8(10.0)\pm 2.0]\times10^{-7}$,
where the value in brackets corresponds to the abundance obtained by using the extinction corrected \SIV\ line flux
(cf. Sect.~\ref{sect:iizw40:extinction}).

Measurements of the oxygen abundance towards \iizw\
\citep[e.g][]{vacca92,pagel92b,walsh93,masegosa94,guseva00,perez03}
give an average value O/H$\sim1/3[$O/H$]_{\sun}$. The value we obtain
for S$^{3+}$/H$^+$ is only about 20\% of the sulphur abundance
expected for a nebula with a metallicity 1/3 of the solar
metallicity. Most of the sulphur must be either in the form of
S$^{++}$ or S$^{4+}$.

Since it has been proven that practically all of the infrared emission
of \iizw\ comes from this compact core \citep[e.g.][]{beck02}, we can estimate the amount of
sulphur locked in the form of S$^{++}$ by using the \SIII\
18.7~\micron\ line flux measured by ISO
\citep[$5.1\pm1.0\times10^{-20}$ W~cm$^{-2}$; cf.][]{verma03}. Adopting the above values for the electron
temperature and density, we obtain $\epsilon_{\rm
SIII\,18.7}=1.28\times10^{-20}$~erg~\cm~s$^{-1}$.
The resulting
S$^{++}$/H$^+$ is $[8.8(10.5)\pm3.0]\times10^{-7}$, where, again, the value in brackets is obtained by adopting the extinction estimated in the previous section.
Hence, S$^{++}$ and S$^{3+}$
account for less than 40\% of the expected sulphur abundance. The rest
must be in the form of S$^{4+}$, which needs energies around 47~eV in
order to be produced. Unfortunately, lines of S$^{4+}$ are not present in the
infrared to prove this result but the line of \OIV\ at 25.9\,\micron\ has been observed by ISO (and the ionisation edge for the creation of this ion is 54.9~eV, even higher).

Larger values of the density will give larger S$^{++}$/H$^+$ and
S$^{3+}$/H$^+$ abundances because collisional de-excitation begins to
play an important role. However, densities much larger than
$10^4$~\cm\ (we note that the critical densities of the \SIII\ 18.7
and \SIV\ 10.5~\micron\ lines are, respectively, $10^4$ and
$3.7\times10^4$~\cm; see \citeauthor{martin:paperii}
\citeyear{martin:paperii}) are necessary in order to produce an
important increase in these abundances. Such high densities are not
probable in \HII\ regions, even in ultracompact ones
\citep[e.g.][]{wood89,kurtz94}.

\subsection{Henize 2-10}
\label{sect:he2-10}

\subsubsection{General description}
\label{sect:he2-10:general}

The radial velocity of \henize\ ($\sim 860-870$~km~s$^{-1}$) yields a
distance anywhere from 6 to 14 Mpc depending on the chosen value of
$H_0$ and assumptions on the Virgocentric flow. As done by other
authors \citep[e.g.][]{vacca92}, we adopt a reasonable compromise of 9~Mpc, which accounts for
Galactic rotation but no Virgocentric flow, and assumes
$H_0=75$~km~s$^{-1}$~Mpc$^{-1}$. At this distance,
1\arcsec\ corresponds to 43.6~pc.

\henize\ was the first emission-line galaxy to exhibit evidence for WR
stars \citep{allen76} and it is considered the prototype for WR
galaxies \citep{conti91}.
%Broad band B and V images \citep{corbin93} reveal that
%\henize\ consists of two distinct starburst regions (designated A and B) at the centre of
%an elliptical stellar envelope about 10 times their size, with a major
%axis diameter of approximately 3.8 kpc ($\sim$1\farcm5).
%These starburst regions, designated A and B, are
%separated by $\sim 8$\arcsec\ and are characterised by nebular
%emission spectra.
%This morphology lead \cite{corbin93} to suggest
%that \henize\ is a single dwarf elliptical experiencing
%``self-induced'' star formation. However, \HI\ and CO aperture
%synthesis observations find evidence for an interaction or merger
%which would have triggered the burst of star formation
%\citep{kobulnicky95}.
%Deep H$\alpha$ images show that \henize\ is
%surrounded by a complex kpc-scale bipolar superbubble centred on
%region A \citep{mendez99}.
%Near-IR emission traces
%the optical and UV \citep{beck97,vacca02} {\bf towards the central nucleus (region A)} and %show an elongated
%ellipse of about 5\arcsec$\times$3\arcsec extended in P.A. 130
%degrees. \cite{beck97} call this structure ``the disc'' and suggest
%that it is the original disc of the galaxy and that accreted CO
%falling onto it is fuelling the burst of star formation.
High-resolution observations in the optical and near-IR \citep[see][]{cabanac05} show a bright central nucleus, generally refered to as region A, resolved into several UV-bright super star clusters which lie in an
arc of about 2\arcsec\ \citep{conti94}. This central nucleus is surrounded by two presumably older star-forming regions, named B and C. Region B, to the east, shows a mixed population of blue and red clusters (only detected at wavelenghts longer than 2.2~\micron). Region C, on the northwest side, has a long tail containing bright red clusters as well.

%The central starburst region A is spatially resolved by HST UV imaging
%\citep{conti94} into several bright pointlike knots which lie in an
%arc of about 2\arcsec. Region B is also resolved into several
%pointlike knots which are scattered randomly throughout the
%regions. These knots seem to be recently formed ($\leq 10$~Myr)
%globular clusters with masses between $10^5$ and $10^6$\,\msun.

VLA radio continuum imaging \citep{kobulnicky99,johnson03} reveals
5 compact $\sim 1$~mJy radio sources in the central region aligned in a
east-west orientation.
These radio sources are labeled 1 to 5 from west to east.
While the global radio spectral index is highly nonthermal ($\alpha
\simeq -0.5$, with $S_{\nu} \propto \nu^{\alpha}$) and consistent with
synchrotron radiation produced in supernova explosions, these compact
radio sources have positive ($\alpha > 0$) spectral indices suggesting
an optically thick thermal bremsstrahlung origin, consistent
with unusually dense \HII\ regions.
%with average electron densities of
%$\sim 10^3-10^4$~\cm\ and radii between 2 ($\sim$0\farcs05) and 4~pc
%($\sim$0\farcs09).
%These radio knots do not show obvious counterparts
%in the near-IR, optical or UV images, indicating that these are the
%youngest, densest and most highly obscured star formation sites in
%\henize.
%They might be the precursors of super-star clusters or
%proto-globular clusters.

   \begin{figure*}[!ht]
   \centering \includegraphics[width=18cm]{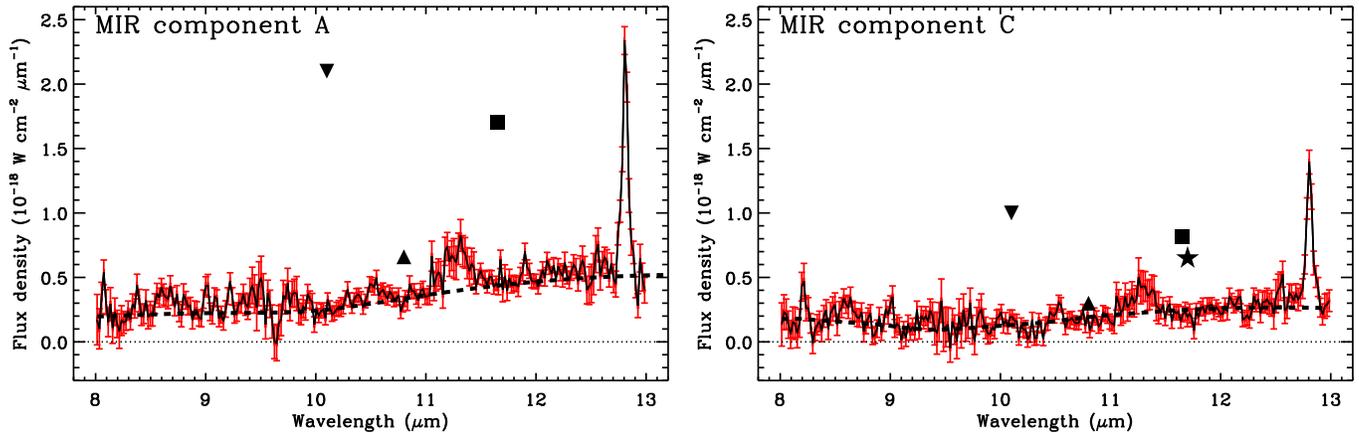}
   \caption{$N$-band spectra of MIR components A (left) and C
   (right) in He\,2--10 with 1$\sigma$ errors.  The spectrum of
   A is roughly twice as bright as that of C.  Both spectra are very similar
   and are characterised by a rising dust continuum, weak PAH emission
   bands at 8.6 and 11.2~\micron\ and a strong \NeII\ 12.8~\micron\
   line. For comparison, we plot the 10.1 and 11.65~\micron\ fluxes
   measured by \cite{sauvage97} for both components A and C (upside
   down triangle and square, respectively). The first one is a broad
   band measurement with a FWHM $\sim 5$~\micron, while the second one
   is a narrow band observation with a FWHM of $\sim 1.7$~\micron.
   The narrow band (FWHM $\sim 1$~\micron) 11.7~\micron\
   flux measured by \cite{beck01} is plotted by a star. In the left
   panel, it coincides with the value quoted by \cite{sauvage97}.  The
   broad band 10.8~\micron\ fluxes measured by \cite{vacca02} for
   radio knots 1+2 (C) and 4 (A) are plotted as a triangle.  The
   dashed line represents the local spline continuum.}
         \label{fig:he2-10}
   \end{figure*}

The presence and importance of dust in the central region was confirmed by high spatial resolution mid-infrared (MIR) images \citep{sauvage97,beck01,vacca02}.
%\henize\ was first imaged in the 10~\micron\ atmospheric window at
%high spatial resolution by \cite{sauvage97}.
The spatial distribution
of the IR emission (which extends less than 5\arcsec) agrees strikingly
with that of the radio continuum.
In fact, the subarsecond
resolution observations by \cite{vacca02} using the Gemini North Telescope
were able to detect 4 of the 5 radio knots observed by
\cite{kobulnicky99}.
%Further high spatial resolution
%mid-infrared images with the Keck I Telescope by \cite{beck01} show 3
%sources, labelled A, B and C (which should not be mistaken by the
%optical components A and B). Source A includes radio knots 4 and 5 by
%\cite{kobulnicky99}, B corresponds to radio knot 3 and C includes
%radio knots 1 and 2.
%The ratio of IR to radio flux for these sources
%confirm that they are compact \HII\ regions rather than supernova
%remnants, whose IR fluxes are many orders of magnitude weaker.
These MIR nebulae provide at least 80\% of the total flux seen by IRAS at
12~\micron\ with a beam that included the entirely galaxy \citep{beck01}.
%The relative fluxes of all 3 components are identical in the radio and IR:
%67\% of the flux is in component A, 8\% in B and 22\% in C.
%\henize\ has been also imaged in the 10~\micron\ region at subarsecond
%resolution using the Gemini North Telescope by \cite{vacca02}. These
%authors were able to detect 4 of the 5 radio knots observed by
%\cite{kobulnicky99}. They have modelled the radio and IR spectra of
%these regions under the assumption of ``scaled-up'' Galactic
%ultracompact \HII\ regions and estimate that the exciting clusters
%have ages $\lesssim 5\times10^6$~Myr and masses greater than about
%$5\times10^5$\,\msun.

Recent high resolution observations in $K_S$ (2.2 \micron), $L'$ (3.8 \micron) and $M'$ (4.8 \micron) bands by \cite{cabanac05} have provided a detailed explanation of the multiwavelength appearance of \henize\ from the optical to the radio. Contrary to what it was thought previously, these authors show that practically all the radio knots can be associated with $K_S$- and $L'$-emitting regions which implies a revision of their physical nature.
%In particular, the positions of radio knots 1+2 and 5 (associated with two MIR nebula) seem to be correlated %to those of two bright $L'$ point sources (L1 and L2) with no counterparts in the optical bands. Radio knot %4, associated to the brightest MIR source, seems to be the counterpart of a group comprising the $L'$ sources %L4b, L4c and L4d, which are easily detectable in the optical wide bands and are quite strong in H$\alpha$. %The brightest $L'$ source, L4a, appears displaced from its possible radio and MIR counterpart, knot 4. Knot %3, on the contrary, does not have a clear association.
These authors tentatively review the classification of the radio knots and only classify knots 1+2 and 5 as bona fide ultradense \HII\ regions. They have counterparts in the NIR but not in the visible which implies a significant optical depth (typically $\gtrsim 10$) and thus a young age although possibly not as young as previoulsy postulated. Regarding knots 3 and 4, they propose that they are supernova remnants mixed with normal \HII\ regions.

\subsubsection{N-band spectrum}
\label{sect:he210:spectrum}

Our MIR observations were centred on the positions of the MIR counterparts to the radio knots 4 and 1+2, named A and C respectively following the nomenclature by \cite{beck01} (see Sect.~\ref{sect:he2-10:general}). These MIR regions A and C should not be mistaken with
the equally named optical components (see previous section).
Fig.~\ref{fig:he2-10} shows the spectra of \henize\ A and C. They show
PAH emission bands at 8.6 and 11.2~\micron\ and the line of \NeII\ at
12.8~\micron, while the presence of silicate in absorption is not clear.
The presence of this feature in absorption is however more  evident in
the MIR spectrum of
\cite{phillips84} obtained within a beam of 5\farcs9.

High spatial observations of \henize\ have been done at 10.1 and
11.65~\micron\ by \cite{sauvage97}, at 11.7~\micron\ by \cite{beck01}
and at 10.8~\micron\ by \cite{vacca02}. The photometric fluxes these
authors have obtained are indicated in Fig.~\ref{fig:he2-10}. While
there is a huge discrepancy between our spectra and the fluxes
obtained by \cite{sauvage97} and \cite{beck01}, they agree rather well
with those of \cite{vacca02}. Possible discrepancies between these
photometric measurements are largely discussed by \cite{vacca02}.
They could be due to differences in the $N$-band filter transmission
profiles and central wavelengths, the use of a wrong colour term in
the photometric calibration, non-photometric conditions during the
observations or to a wrong background correction beneath the knots.

In the case of \henize\ A, we have compared the spatial
distribution of the \NeII\ and 11.2~\micron\ peaks with that of the
continuum at 12~\micron. No differences are evident, consistent with the
fact that the emission is only slightly resolved.

\subsubsection{Line fluxes}
\label{sect:he210:lines}

We detect the \NeII\ line towards \henize\ A and C.  \NeII\ line
fluxes and upper limits of the non-detected lines are measured as
described in Sect.~\ref{sect:ngc3256:lines} and listed in
Table~\ref{table:fluxes:he2-10}.

\begin{table}[!ht]
\caption{Line fluxes and $3\sigma$ upper limits measured towards \henize. These are
	compared with fluxes measured using other apertures/slits.}
  \label{table:fluxes:he2-10}
  \begin{center}
    \leavevmode
    \begin{tabular}[h]{lcccccc}
      \hline \hline \\[-7pt]
    \multicolumn{1}{c}{Line} &
    \multicolumn{1}{c}{$\lambda$} &
    \multicolumn{1}{c}{\henize\ A} &
    \multicolumn{1}{c}{\henize\ C} &
    \multicolumn{1}{c}{Aperture} &
    \multicolumn{1}{c}{Ref.} \\
    \multicolumn{1}{c}{} &
    \multicolumn{1}{c}{(\micron)} &
    \multicolumn{2}{c}{(10$^{-20}$ W~cm$^{-2}$)} &
    \multicolumn{1}{c}{(\arcsec)} &
    \multicolumn{1}{c}{} \\[5pt] \hline \\[-7pt]

\ArIII &  9.0 & $< 1.8$    & $< 1.6$       & 1.2 & 1\\
\SIV   & 10.5 & $< 1.1$    & $< 1.3$       & 1.2 & 1\\
\NeII  & 12.8 & $12 \pm 1^\star$ & $6.8 \pm 0.5$ & 1.2 & 1\\[5pt]

\NeII  & 12.8 & \multicolumn{2}{c}{$19.5 \pm 3.5$} & 5.9 & 2\\
\NeII  & 12.8 & \multicolumn{2}{c}{20}             & 5.9 & 3\\[5pt]

\ArIII &  9.0 & $1.6 \pm 0.5$ & -- & 2.0 EW & 4\\
\SIV   & 10.5 & $0.5 \pm 0.3$ & -- & 1.6 EW & 4\\
\NeII  & 12.8 & $20 \pm 2$    & -- & 2.0 NS & 4\\
\NeII  & 12.8 & $25 \pm 5$    & -- & 1.6 EW & 4\\[5pt]
\hline

    \end{tabular}
  \end{center}
$^\star$ When correcting for slit losses, we obtain a total line flux of about $(16-20)\times10^{-20}$~W~cm$^{-2}$ (cf. Sect.~\ref{sect:he210:lines}).\\
REFERENCES:
(1) This work;
(2) \cite{phillips84};
(3) \cite{roche91};
(4) \cite{beck97}.
\end{table}

\begin{table}[!ht]
\caption{Continuum subtracted 8.6 and 11.2~\micron\ PAH band fluxes (in units of
        $10^{-20}$ W cm$^{-2}$) measured towards \henize\ A and C.}
  \label{table:pahs:he210}
  \begin{center}
    \leavevmode
    \begin{tabular}[h]{lcccc}
      \hline \hline \\[-7pt]
    \multicolumn{1}{c}{Component} &
    \multicolumn{1}{c}{8.6 \micron} &
    \multicolumn{1}{c}{11.2 \micron} &
    \multicolumn{1}{c}{Aperture (\arcsec)} &
    \multicolumn{1}{c}{Ref.}
    \\[5pt] \hline \\[-7pt]
A          & $< 5^\triangle$ & $9 \pm 5$ & 1.2 & 1\\
C          & $< 7^\triangle$ & $6 \pm 5$ & 1.2 & 1\\
? & --    & $12 \pm 2^\natural$ & 5.9 & 2\\[5pt]

  \hline

   \end{tabular}
  \end{center}
$^\triangle$ Given as an upper limit due to the large error bars
(see Fig.~\ref{fig:he2-10}).
$^\natural$ The central position of this observation is uncertain. It is likely centred on
component A.\\
REFERENCES:
(1) This work;
(2) \cite{phillips84}.
\end{table}

%XANDER {\bf why are these 8.6um fluxes upper limits?}

Regarding \henize, spectra of its nuclear region have been previously
obtained by \cite{phillips84}, \cite{roche91} and \cite{beck97}. A
comparison between the fluxes obtained by these authors and ours is
shown in Table~\ref{table:fluxes:he2-10}. The \ArIII\ and \SIV\ line
fluxes obtained by \cite{beck97} towards \henize\ A agree well with
our upper limits. Their \NeII\ line fluxes are, however, larger than
the value we get. Still, slit losses might be affecting the \NeII\
line flux we measure. In Sect.~\ref{sect:he210:spectrum}, we estimate
that our slit might only be registering about 60-75\% of the total
flux emitted by component A. Considering this loss, the total \NeII\
line flux would be about $(16-20)\times10^{-20}$~W~cm$^{-2}$, a value
that perfectly agrees with the fluxes obtained by \citeauthor{beck97}
using larger slit widths. The comparison with the \NeII\ flux obtained
by \cite{phillips84} and \cite{roche91} is more complicated since they
do not specify the central position of their apertures. However, it
seems natural to assume that their apertures are centred on the
brightest component A. Their \NeII\ fluxes would then agree well with
the \NeII\ fluxes obtained by \citeauthor{beck97} towards A and our
value when corrected for slit losses.

\subsubsection{PAH bands}

PAH emission bands at 8.6 and 11.2~\micron, although very weak, have
been detected in the spectra of \henize\ A and C. The PAH fluxes are
determined by subtracting a local spline continuum
(Fig.~\ref{fig:he2-10}) and are listed in
Table~\ref{table:pahs:he210}. The continuum subtracted spectra of both
components are identical within the errors except for the \NeII\
line. The 11.2~\micron\ PAH flux has been previously measured towards the
central 5\farcs9 of \henize\ by \cite{phillips84}. They quote a PAH
band flux which is in reasonably agreement with the combined flux we
obtain for components A and C.

\subsubsection{Extinction}
\label{sect:he2-10:extinction}

%XANDER {\bf What? Silicate absorption?}

The central region of \henize\ is a complex one and hence is not surprising that extinctions
measured at different wavelengths do not agree well with one another.
For instance, \cite{vacca92} measure $A_{\rm v} \sim 1.7$ from optical spectroscopy.
\cite{cabanac05} derive $A_{\rm v}=1.25$ from the H$_\alpha$/Br$\gamma$ ratio and $A_{\rm v}=10.5$ from the Br$\gamma$/Br10 ratio.
From the
Br$\alpha$/Br$\gamma$ line ratio observed by \cite{kawara89} one finds as well an extinction
of $A_{\rm v} \sim 10$~mag which agrees well with the $15\pm5$~mag
obtained by \cite{phillips84} from the depth of the silicate
absorption. On the other hand, the beam-averaged CO column density towards the large
optical starburst A implies an extinction of $\sim 30$~mag.

As mentioned in previous sections, the extrapolation of the extinction
in the optical and near-IR to the MIR regime is not
direct. Moreover, as in the case of NGC\,3256
(cf. Sect.~\ref{sect:ngc3256:extinction}), the presence of PAH bands
hampers the fitting of the dust continuum. Therefore, as it has been
done previously, we will simply give rough estimates of the extinction
in the MIR by using the ``astronomical silicate'' with $A_{\rm V}/A_{\rm sil}=18.5$
(see Sect.~\ref{sect:ngc3256:extinction})
and considering
$A_{\rm V}=10$. Using this, we have $A_{12.8}=0.33A_{\rm
sil}=0.2$~mag. When applying this extinction value to the \NeII\ line
fluxes measured for components A and C, we obtain, respectively, $\sim
14\times10^{-20}$ and $\sim 8\times10^{-20}$~W~cm$^{-2}$.

\subsubsection{Ionic abundances}
\label{sect:ab:he2-10}

As it has been previously done in Sect.~\ref{sect:ab:ngc3256},
we can estimate the ionic abundance of Ne$^+$ with respect to H$^+$.

 Measurements of the Br$\gamma$ line flux have only been obtained at low spatial resolutions, with apertures of, e.g. 7\farcs1$\times$3\farcs5 \citep{kawara89}, 5\arcsec \citep{doyon92} or 2\farcs4$\times$15\farcs6 \citep{vanzi97}.
Nevertheless, we can give estimates
of the Br$\gamma$ emission of knots A and C from the
associated radio emission.

\cite{johnson03} have recently provided 0.7~cm flux densities. Measurements at this frequency are likely to
contain an insignificant non-thermal contribution and are usually
optically thin. They measure flux densities of 2.91~mJy for component
A (equivalent to radio source \#4) and 1.87~mJy for component C (which
results by adding the contributions of radio sources \#1 and
\#2). Adopting an electron temperature of 10000~K
\citep{johansson87,vacca92} and the electron densities listed in
Table~\ref{table:prop}, we estimate Br$\gamma$ fluxes of
$0.41\times10^{-20}$ and $0.26\times10^{-20}$~W~cm$^{-2}$ for components A
and C (we will consider a standard 20\% uncertainty for these line fluxes).
These values can be compared with the previously mentioned Br$\gamma$ observations. These
authors find Br$\gamma$ fluxes in between $4.4\times 10^{-21}$ and $6.5\times 10^{-21}$~W~cm$^{-2}$ which, adopting $A_{\rm V}=10$
($A_{\rm K}=1.1$, \citeauthor{rieke85} \citeyear{rieke85}) give unreddeded fluxes of
$(1.2-1.8)\times10^{-20}$~W~cm$^{-2}$, about 2--3 times larger than the combined
value we obtain for the MIR components A and C using the associated radio free-free emission.

The emission coefficient of the Br$\gamma$ line for \tel$=10^4$~K is
$\epsilon_{\rm Br\gamma}=5.98\times10^{-27}$~erg~\cm~s$^{-1}$.  In the
case of component A, with \tel$=10^4$~K and \den$=4290$~\cm, the
emission coefficient of the \NeII\ line is $\epsilon_{\rm
[NeII]\,12.8}=8.38\times10^{-22}$~erg~\cm~s$^{-1}$, while in the case
of component C, with \tel$=10^4$~K and \den$=2240$~\cm, $\epsilon_{\rm
[NeII]\,12.8}=8.42\times10^{-22}$~erg~\cm~s$^{-1}$.
Following
Eq.~\ref{eq:ab}, we obtain Ne$^+$/H$^+$(A)$\gtrsim[1.5(1.8)\pm 0.4]\times10^{-4}$
and Ne$^+$/H$^+$(B)$\gtrsim[1.6(1.9)\pm 0.4]\times10^{-4}$, where the values in
bracktets are obtained by correcting the \NeII\ fluxes from extinction
(Sect.~\ref{sect:he2-10:extinction}).
These abundances give a lower limit to the elemental
abundance of Ne of $(1-2)$[Ne/H]$_{\sun}$.

Measurements of the oxygen abundance towards the optical starburst A
where these bright infrared knots are located (see
Sect.~\ref{sect:he2-10:general}) based on long-slit optical spectra
give values between $\sim 0.5$
and $\sim 0.8$ times the solar
oxygen abundance \citep{johansson87,vacca92,kobulnicky95}. These values have been
recomputed using the P-method of
\cite{pilyugin01} for high metallicity regions and considering
$12+\log[{\rm O/H}]_{\sun}=8.69$ \citep{allende01}.
%The lower limit we obtain for the Ne abundance indicates that these
%knots have at least a solar metal content. The same result is found by
%\cite{beck01}, who obtain a Ne abundance around twice solar from the
%\NeII\ flux observed by \cite{beck97} and the 2~cm observations by
%\cite{kobulnicky99}. This might be the effect of a general metal
%enrichment or of an enhancement of neon with respect to oxygen.
Our lower limits for the Ne/H abundance and the value estimated
by \cite{beck01} from similar MIR and radio observations (around twice solar)
tend to indicate an abundance ratio of solar or higher.
Taken at face value, this might be the effect of a metal enrichment
in the infrared knots or of an enhancement of neon.
More likely, this apparent discrepancy is due to an underestimate
of the true solar Ne abundance, as recently suggested from
helioseismological and stellar studies. Indeed, our estimated
Ne/H abundance is subsolar and hence reconciled with a subsolar O/H value
if we adopt the latest upward revisions of the solar Ne abundance
suggested from helioseismology \citep{antia05,bahcall05} and from solar type stars
\citep{drake05}.  Moreover, we might be underestimating the Br$\gamma$ line fluxes associated with the MIR components as suggested by the comparison with direct measurements of Br$\gamma$.

\section{Discussion}
\label{sect:discussion}

\subsection{Importance of high spatial resolution observations}
\label{sect:iso}

\begin{figure}[!ht]
   \centering \includegraphics[width=8.5cm]{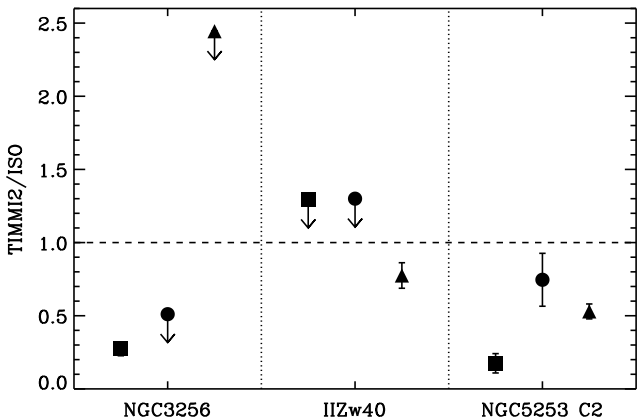}
   \caption{Comparison between the line fluxes measured by TIMMI2 and
        ISO. Symbols correspond to \NeII\ 12.8~\micron\ (square),
        \ArIII\ 9.0~\micron\ (circle) and \SIV\ 10.5~\micron\ (triangle).
        The lines are ordered by increasing ionisation potential: 21.6,
        27.6 and 34.8~eV respectively.
        In the case of NGC\,3256, we have added the
        contributions of the N and S nuclei to the TIMMI2 flux since
	the ISO aperture includes
	both sources. The
        dashed line represents the one-to-one relation.}
         \label{fig:iso}
   \end{figure}

Figure~\ref{fig:iso} shows a comparison between the fluxes measured by
TIMMI2 and ISO for NGC\,3256, \iizw\ and NGC\,5253~C2. In the case of
NGC\,3256 (cf. Sect.~\ref{sect:ngc3256:lines}), only 30\% of the
\ArIII\ and \NeII\ fluxes comes from the two galactic nuclei. In the
case of NGC\,5253~C2, only 20\% of the
\NeII\ flux measured by ISO comes from this super-star cluster, while
the percentages are larger for the \SIV\ ($\sim 50$\%) and \ArIII\
($\sim 80$\%). In our previous work, where we presented the TIMMI2
spectrum of the IR bright nebula C2 in NGC5253
\citep[cf.][]{martin:ngc5253}, we showed the implications that this
has for the interpretation of line fluxes in terms of the properties
(age, IMF, etc.) of the ionising cluster. Only in the case of \iizw,
there seems to be a good correspondence between ISO and our high
spatial resolution observations, suggesting that practically  most (if
not all) of the MIR fluxes of the galaxy is confined to the compact
infrared source.

%{\bf Hence, Fig.~\ref{fig:iso} shows the importance of high spatial resolution observations
%of extra-galactic sources. Only such observations provide accurate
%measurements of the line and molecular band fluxes emitted by the
%cluster or galactic nucleus.}
%XANDER {\bf this point isn't clear from the
%previous discussion. See below. (remove this upper lines.)}

\subsection{A comparison with the infrared supernebula C2 in NGC\,5253}

In Table~\ref{table:prop} we show a summary of the properties (size,
electron density, $Q_0$, metallicity, presence of PAHs and excitation)
of the sources described in this work. They are compared with the
properties of the embedded super-star cluster C2 in NGC\,5253
\citep{martin:ngc5253}.

\begin{table*}[!ht]
\caption{Summary of properties of the different infrared regions observed in NGC\,3256 \iizw, \henize\ and NGC\,5253.}
  \label{table:prop}
  \begin{center}
    \leavevmode
    \begin{tabular}[h]{lrrrrrr}
      \hline \hline \\[-7pt]
    \multicolumn{1}{c}{Property} &
    \multicolumn{1}{c}{NGC\,3256~N} &
    \multicolumn{1}{c}{NGC\,3256~S} &
    \multicolumn{1}{c}{\iizw} &
    \multicolumn{1}{c}{\henize\ A} &
    \multicolumn{1}{c}{\henize\ C} &
    \multicolumn{1}{c}{NGC\,5253 C2}
    \\[5pt] \hline \\[-7pt]

$D$ (Mpc)$^{\rm a}$        & 37      & 37      & 9.2     & 9.0        & 9.0        & 3.3$^{(13)}$ \\
Diameter (pc)$^{\rm b}$    & 316$^{(1)}$ & $<179^{(6)}$&$<22.3^{(7)}$& 5.8$^{(10)}$ & 8.4$^{(10)}$ & 1.6$^{(13)}$ \\
$n_{\rm e}$ (cm$^{-3}$)$^{\rm c}$ & 1000$^{(2)}$& --      &$>1700^{(8)}$& 4290$^{(10)}$& 2240$^{(10)}$& 50000$^{(13)}$\\
$Q_0$ ($10^{52}$ s$^{-1}$)$^{\rm d}$ & 60$^{(3)}$  & 15$^{(3)}$  & 10$^{(8)}$  & 2.6$^{(11)}$ & 1.7$^{(11)}$ & 2$^{(13)}$ \\
$Z/Z_{\sun}$               & 1.3$^{(4)}$ & 1.3$^{(4)}$ & 1/3$^{(9)}$ & $>1.2^{(12)}$ & $>1.2^{(12)}$ & 1/3$^{(13)}$ \\
$A_{\rm V}$                & 5.6$^{(14)}$ & 10.2$^{(14)}$ & $10^{(15)}$ & 10$^{(16)}$ & 10$^{(16)}$ &
$17\pm6^{(17)}$ \\
PAH bands                  & yes     & yes     & no      & yes        & yes        & no$^{(13)}$ \\
(\SIV/\NeII)$_{\rm TIMMI2}$ $^{\rm e}$ & $<0.05$ & $<0.40$ & $>6.7$ & $<0.10$ & $<0.21$ & $12.9\pm4.6^{(13)}$ \\
(\SIV/\NeII)$_{\rm ISO}$ $^{\rm f}$ &\multicolumn{2}{c}{$0.010\pm0.002^{(5,18)}$}&$12\pm2^{(5)}$ &-- & --& $5.6\pm0.8^{(5)}$\\[5pt]
  \hline

   \end{tabular}
  \end{center}
$^{\rm a}$ Distance to the source.
$^{\rm b}$ Source diameter.
$^{\rm c}$ Electron density.
$^{\rm d}$ Ionising photon luminosity.
$^{\rm e}$ \SIV/\NeII\ line ratio observed with TIMMI2.
$^{\rm f}$ \SIV/\NeII\ line ratio observed with ISO.
$^1$ \cite{siebenmorgen04} has recently obtained an image at 10.4~\micron\
with a resolved (FWHM) diameter of 1\farcs76.
$^2$ Estimated from the [\ion{S}{ii}] $\lambda \lambda$6717/6731 line ratio of 0.96 measured
by \cite{lipari00}.
$^3$ Estimated from the extinction-corrected Br$\gamma$ line flux obtained by
\cite{kotilainen96}, see Sect.~\ref{sect:ab:ngc3256}.
$^4$ From \cite{lipari00} assuming $12+\log$(O/H)$_{\odot} = 8.69$ \citep{allende01}.
$^5$ From \cite{verma03}.
$^6$ Upper limit given by our spectroscopic observations (cf. Sect.~\ref{sect:ngc3256:spectrum}).
$^7$ From the $N$-band image by \cite{beck02}.
$^8$ Estimated from the 2~cm observations by \cite{beck02} assuming $T_{\rm e}=12500$~K, see Sect.~\ref{sect:ab:iizw40}.
$^9$ Average value of different estimates
\citep{vacca92,pagel92b,walsh93,masegosa94,guseva00,perez03}. It assumes $12+\log$(O/H)$_{\odot} = 8.69$ \citep{allende01}.
$^{10}$ From \cite{vacca02}.
$^{11}$ Estimate given by \cite{johnson03}.
$^{12}$ See Sect.~\ref{sect:ab:he2-10}.
$^{13}$ From \cite{martin:ngc5253}.
$^{14}$ See Sect.~\ref{sect:ngc3256:extinction}.
$^{15}$ Sect.~\ref{sect:iizw40:extinction}.
$^{16}$ See Sect.~\ref{sect:he2-10:extinction}.
$^{17}$ From \cite{martin:ngc5253}.
$^{18}$ The ISO aperture includes both N and S nuclei.
\end{table*}

Overall the properties compiled in Table~\ref{table:prop} are quite
suggestive of the regions/``nuclei"
of NGC 3256 as complex regions, possibly hosting several separate star forming
regions. Possibly this is also the case for the remaining objects, as
it seems quite improbable that regions as extreme as C2 of NGC\,5253 are
sufficiently
common. Indeed, there are already indications, mostly from radio observations,
that \iizw\ and regions A and B of \henize\ consist over several smaller
star forming regions, or super-star complexes \cite[e.g.][]{beck02,johnson05},
on spatial scales smaller than the ones achieved here.

Hence, the two nuclei in NGC\,3256 are characterised by their large sizes and
values of $Q_0$, which indicate that they may probably be ionised by multiple
clusters
having $6\times10^4$ and $1.5\times10^4$ O7 equivalent stars,
respectively. Their MIR spectra show PAHs and their excitation
(measured by the \SIV/\NeII\ line ratio) is relatively low.

More interesting is the comparison between the compact objects in
\iizw\ and \henize\ with the embedded supernebula C2 in NGC\,5253. The
optically thick thermal radio emission and compactness of these
sources have lead to assume them to be ``scaled-up'' ultracompact
\HII\ regions excited by super-star clusters which have recently
formed \citep[e.g.][]{vacca02}.
%C2 and the IR sources in \henize\ seem to be heavily embedded,
%with visual extinctions of the order of 17 mag
%\citep[cf. Sect.~\ref{sect:he2-10:general};][]{martin:ngc5253}. However,
%their MIR spectra are completely different.

The spectrum of C2 is
characterised by a strong \SIV\ line and does not show PAHs. On the
contrary, the spectra of the MIR components A and C in \henize\ show a strong \NeII\ line
and have PAHs. The difference between their spectra could be due to the
different hardness of the radiation field, as traced by the
\SIV/\NeII\ line ratio and which is directly influenced by the
metallicity, among other factors.
While the IR sources in \henize\ might have a solar (or
supersolar) metallicity and have a low \SIV/\NeII\ line ratio, the opposite
is true for C2.
Even more, it is quite likely that cluster C2 in NGC\,5253 is quite different from
regions A or C in \henize\ in the sense that it may be much more embedded: it seems that this cluster has no optical counterpart \citep[e.g.][]{alonso04}. Hence, it is not surprising that their IR spectrum is different.

In terms of metal content and excitation, C2 is more
similar to the compact IR nebula in \iizw. However, the nebula in
\iizw\ is not so deeply embedded (visual extinction is of the order
of 2-3 mag, although some authors give values as large as 10 mag; see
Sect.~\ref{sect:iizw40:extinction}) and is much less compact than C2.

\subsection{PAH emission}

\subsubsection{The ground-based MIR/FIR diagnostic diagram for star forming
regions}

\citet{peeters04} have examined the use of PAH bands as tracers of
star formation. In order to distinguish the different natures of the
galaxies, i.e. AGN-dominated, starburst-dominated or heavily obscured,
these authors present a new MIR/FIR diagnostic, along the lines
of previous works by e.g. \citet{genzel98}, \citet{Laurent00} and
\citet{clavel00}, based on the ratio of the 6.2\,\micron\ PAH
emission band to FIR flux and the ratio of the 6.2\,\micron\ continuum
to FIR flux. These two ratios isolate the strongly obscured galaxies
while the 6.2\,PAH-to-continuum ratio provides a very clear handle on
any AGN contribution to the MIR. The comparison with Galactic objects reveals that
most of the Ultra-Luminous IR Galaxies (ULIRGs) are found co-located with compact \HII\
regions in contrast to normal and starburst galaxies, which are mainly
co-located with exposed PDRs.
%
%This diagnostic is also applied to a Galactic sample.
%
%Both ratios vary clearly within their sample of
%\HII\ regions and this range extends up to the reflection nebulae (RNe)
%and the (diffuse)
%ISM lines of sight. In addition, the 6.2\,PAH-to-continuum ratio
%provides a very clear handle on any AGN contribution to the
%MIR. Indeed, AGNs are found to segregate in two groups; most
%Seyfert2's are located with the normal and starburst galaxies, while
%most Seyfert1's show strong 6.2\,cont/FIR ratios. This diagram further
%reveals the spectral resemblance of starburst and normal galaxies to
%exposed PDRs rather than (slightly embedded) compact \HII\
%regions. Ultra-luminous IR galaxies (ULIRGs) show a diverse spectral
%appearance. Some show a
%typical AGN hot dust continuum. More, however, are either
%starburst-like or show signs of strong dust obscuration in the
%nucleus.

A problem with this and other diagrams is that they are not
accessible to ground-based observations. We have no access to the
6.2 or 7.7\,\micron\ PAH bands through the atmospheric MIR window
($\sim8-13$~\micron) and the same difficulty arises in the cases of
the diagnostics by \citet{genzel98} and \citet{Laurent00}, which are also
based on lines such as \OIV\ at 25.9~\micron\ or the ratio of warm
(14--15~\micron) to hot (5.1--6.8~\micron) dust continuum.
%The extra-galactic nuclei and embedded star clusters discussed in this
%work and in the paper by \citet{siebenmorgen04} further extend the
%sample included by \citet{peeters04} to other extra-galactic environments
%of star formation activity and hence can shed more light on this
%issue. Unfortunately, we have no access to the 6.2\,\micron\ PAH band
%through the atmospheric MIR window ($\sim8-13$~\micron).
Therefore, we have adapted the MIR/FIR diagnostic diagram presented in
\citet{peeters04} in order to include the 11.2\,\micron\ PAH band and
the 11.2\,\micron\ continuum instead and test whether similar
conclusions are reached
compared to previous works based upon the 6.2 or 7.7 \micron\ PAH
bands.

%the extra-galactic nuclei and embedded star clusters
%discussed in this work and in the paper by \citet{siebenmorgen04} further extend the
%sample included by \citet{peeters04} to other extra-galactic environments
%of star formation activity.

\begin{figure*}[!t]
   \centering \includegraphics[width=9cm, angle=90]{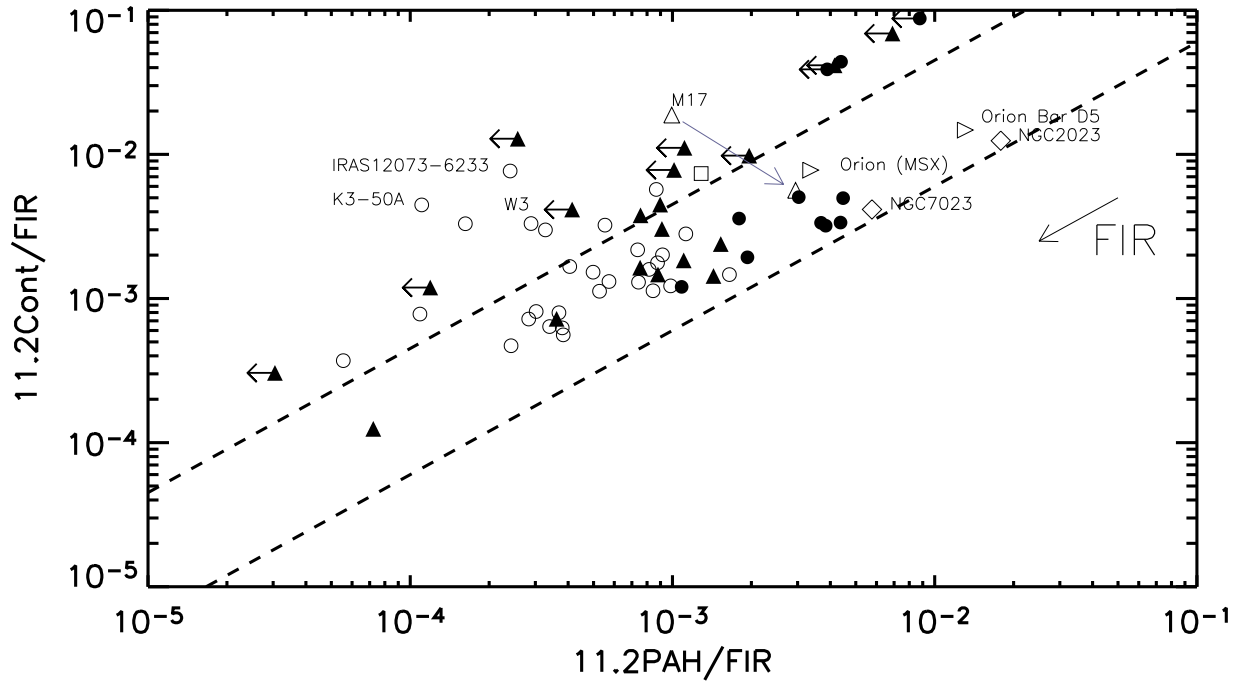}
   \caption{Ground-based MIR/FIR diagnostic diagram for star forming
regions. Galactic \HII\ regions are presented by open circles, Orion
by right triangles, the extended \HII\ region M17 by open triangles,
reflection nebulae (RNe) by open diamonds, 30\,Doradus by an open
square, TIMMI2 observations of extra-galactic nuclei and star clusters
by filled triangles \citep[this work and][]{siebenmorgen04} and
ISOPHOT observations of extra-galactic nuclei by filled circles
\citep{siebenmorgen04}. The dotted lines correspond to a
11.2\,PAH/continuum ratio of 0.22 and 1.66 from top to bottom.  The
effect of an overestimate of the far-IR (FIR) flux by a factor of 2 is
indicated by an arrow parallel to the dotted lines. The second arrow
indicates the shift in position from the pointing inside the \HII\
region towards that of the molecular cloud in the extended star
forming region M17.  This diagram is being adapted from that of
\citet{peeters04} using the 11.2\,\micron\ PAH instead of the
6.2\,\micron\ PAH. }
   \label{fig:dustvspah}
   \end{figure*}

The use of the 11.2\,\micron\ PAH band has however clear consequences
on the interpretation.  The PAH emission band and the dust continuum
at this wavelength are likely to trace different carriers due to a
difference in temperatures, size, heating mechanism and charge state
in case of the PAHs.  To first order, the fraction of total PAH flux
emitted in the 11.2\,\micron\ PAH band varies only from 9 to 27\% with
an average of 18$\pm$5\% and that emitted in the 6.2 \,\micron\ PAH
band varies only from 14 to 38\% with an average of 28$\pm$4\%
\citep{Peeters02}. Given the order of variation observed in the
6.2\,cont/FIR \citep{peeters04} and 11.2\,cont/FIR (see
Fig.~\ref{fig:dustvspah}), both ratios are representative of the
PAH/FIR ratio. However, this clearly does not hold for the dust
continuum emission.

The 11.2\,MIR/FIR diagnostic diagram is shown in
Fig.~\ref{fig:dustvspah}. Note that the FIR fluxes of the
extra-galactic sources have been calculated from the IRAS fluxes and
except for NGC\,3256 N and S, no correction has been made to account
for the aperture difference.  As a consequence, the FIR fluxes
presented here are likely overestimated and the use of an ``aperture
corrected'' FIR flux will move the sources in the opposite direction
to that indicated by the FIR arrow shown on the plot.  However, since
most of the observed extra-galactic nuclei and star clusters are
probably the dominating source of the FIR flux, this correction is
expected to have little influence on the results we discuss here.

The sample of Galactic sources shows a general correlation of the
strength of the 11.2\,PAH/FIR with the 11.2\,cont/FIR.  This is very
similar to that seen for the MIR/FIR diagram based upon the
6.2\,\micron\ band, as discussed by \citet{peeters04}. This general
correlation reveals that the 11.2\,PAH-to-continuum ratio
is relatively constant in the full sample while
%
%their strength relation to the total FIR dust emission
%
the 11.2PAH/FIR ratio changes by almost three orders
of magnitude.  Typically, the 11.2\,PAH-to-continuum ratio is very low
for deeply embedded ultra-compact \HII\ regions (like e.g. W3, K3-50A)
and increases when the characteristics of the region changes to that
of an exposed PDR such as Orion and M17.

As shown by Fig.~\ref{fig:dustvspah}, the galaxies observed with
ISOPHOT by \citet[][]{siebenmorgen04}, represented by filled circles,
are located close
to the exposed PDRs such as Orion and M17. A similar result was
found by \citet{peeters04} for ISOPHOT observations of normal and
starburst galaxies in the 6.2\,MIR/FIR diagram. In contrast, galaxies
observed with TIMMI2 (filled triangles) are found to have a lower
11.2\,PAH/FIR ratio and thus to be located nearer to the
Galactic \HII\ regions. For the 6.2\,MIR/FIR diagram, this is also
observed for the ULIRGs \citep{peeters04}.  This difference between
the ISOPHOT and TIMMI observations is certainly due to the aperture
difference between the ISOPHOT, TIMMI2 and IRAS data (the latter for
the FIR determination). The much larger ISOPHOT aperture ($\sim$
20\arcsec) likely includes much of the surrounding PDRs while the
TIMMI2 slit measures mainly the central emission of the \HII\ region
or nucleus although it does cross the ionisation front
and the PDR where the PAH emission originates from
\citep[e.g.][]{Sellgren:90, Tielens:93, Tielens:anatomyorionbar:93,
Verstraete:m17:96}. This is similar to the results obtained by
\cite{siebenmorgen04} that compare the ISOPHOT data and TIMMI2 spectra of
galaxies. In the 6.2\,MIR/FIR diagram, the ISOPHOT
observations of normal galaxies and starburst galaxies are also found
close to the location of exposed PDRs and are not co-located with the
Galactic \HII\ regions \citep{peeters04}. This suggests that the PAH
emission is probably diffuse and extended and thus difficult to detect
by TIMMI2. Consequently, the PAH emission rather seems to trace
exposed PDRs and the ISM of galaxies (as seen by ISOPHOT) than the
dense star forming regions (as seen by TIMMI2).

   \begin{figure}[!t]
   \centering \includegraphics[width=8cm]{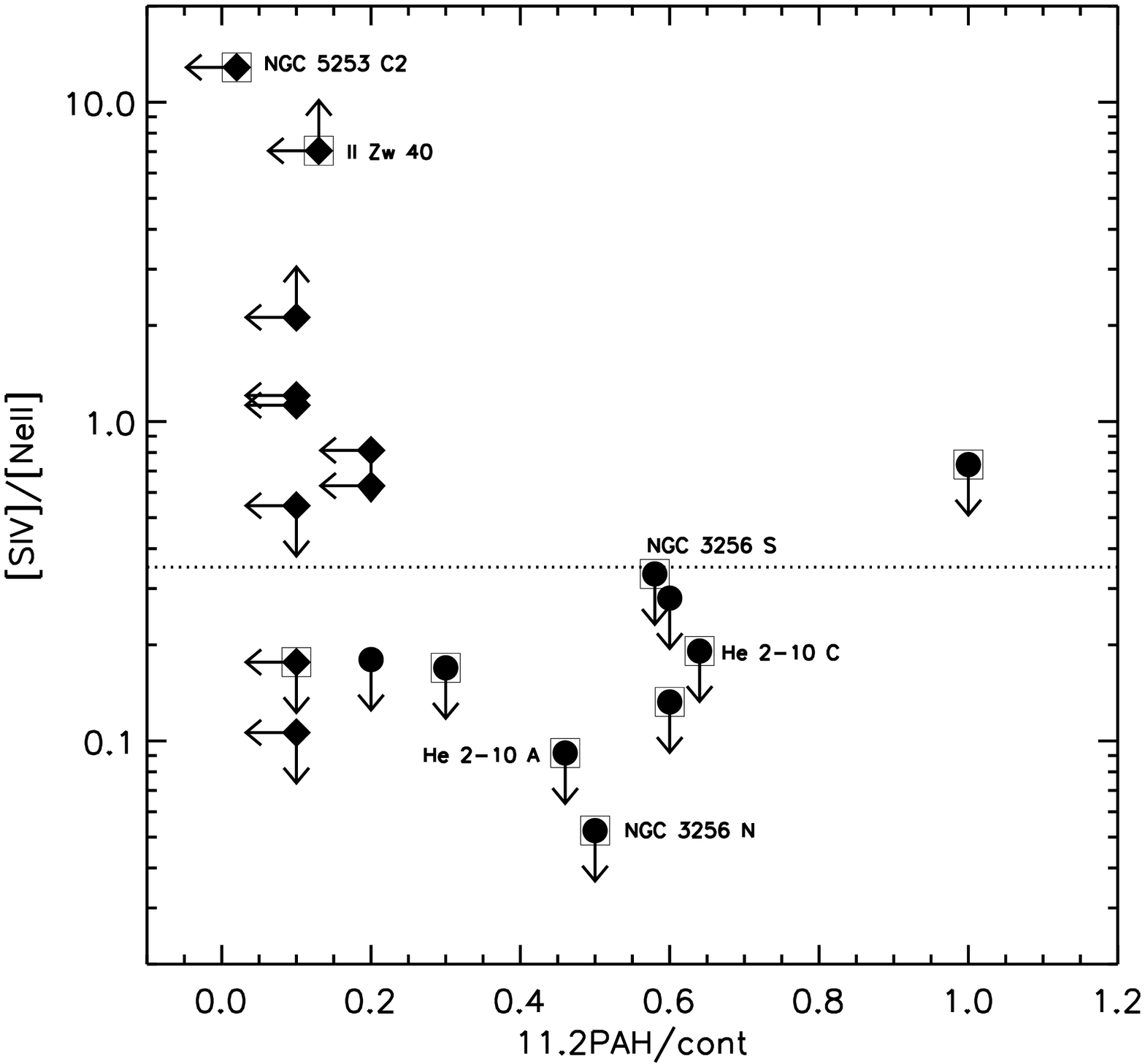}
   \caption{Comparison of the
   \SIV\,10.5\,\micron/\NeII\,12.8\,\micron\ line ratio with the
   11.2\,\micron\,PAH/continuum ratio for extra-galactic
   nuclei and embedded super-star clusters. We combine the sources in
   Table~\ref{table:prop} with the galactic nuclei observed by
   \cite{siebenmorgen04} using also TIMMI2.  Sources plotted as
   circles show PAH bands in their MIR spectra and are concentrated
   predominantly below \SIV/\NeII$\sim 0.35$. Sources plotted as
   diamonds do not show PAHs and have mostly line ratios above this
   limit of 0.35. Nuclei/star clusters in starburst galaxies are
   marked by a square. The other sources are nuclei in Seyfert
   galaxies.}
        \label{fig:corr1}
   \end{figure}

A large number of our sources show no evidence of PAH emission. In
that regard, note that the extra-galactic sources where PAHs are
detected are situated between the two dotted lines in
Fig.~\ref{fig:dustvspah}
%
% (which correspond to a 11.2\,PAH-to-continuum
%ratio of 1.66 and 0.22 from top to bottom)
%
while those without PAH
detections are found above the top dotted line, indicating that in
these objects the 11.2\,PAH-to-continuum ratio is lower.  This is not
influenced by an overestimate of the FIR flux since a lower FIR flux
would move the sources in the opposite direction of the FIR arrow,
parallel to the dotted lines; neither it is influenced by the fact
that we are dealing with upper limits. In fact, if the upper limit
becomes more stringent, the source will move horizontally towards the
left, giving an even lower 11.2\,PAH-to-continuum ratio. Together with
these extra-galactic sources, some Galactic \HII\ regions also fall
above these dotted lines, all of which are ultra-compact \HII\ regions
which have very strong radiation fields and MIR dust continuum.

Concluding, we have constructed a new MIR/FIR diagnostic diagram based on
the 11.2\,\micron\ PAH band and
the 11.2\,\micron\ continuumm and proven
that the same conclusions are reached when
compared to previous works based upon the 6.2 or 7.7 \micron\ PAH bands. This is useful since
the 11.2\,\micron\ PAH band is easily accessible from ground-based observations.

\subsubsection{The \SIV/\NeII\ line ratio and the survival of PAHs}

We analyse now the effect of the hardness of the radiation field,
i.e. the survival of PAHs in environments with energetic photons. We
have a measure of the hardness of the radiation field through the
\SIV/\NeII\ line ratio. Assuming that all sources have the same S/Ne
abundance ratio, variations in \SIV/\NeII\ probe the hardness of the
radiation field at energies above $\sim22$~eV.
Fig.~\ref{fig:corr1} compares \SIV/\NeII\
with the 11.2\,PAH-to-continuum ratio for the objects analysed in this work
(cf. Table~\ref{table:prop}) and those observed by
\cite{siebenmorgen04} using also TIMMI2. This figure clearly shows
that sources with PAHs have a \SIV/\NeII\ ratio $\lesssim 0.35$, while
sources which do not show PAHs predominantly have \SIV/\NeII\ ratios
above 0.35. Exceptions to this behaviour are Centaurus~A and M83,
which do not show PAHs but still have a \SIV/\NeII\ ratio lower than
0.35.  Centaurus~A is the closest example of an AGN. While the
spectrum of this galaxy obtained through large ($>20$\arcsec)
apertures is PAH dominated, the 3\arcsec\ spectrum by
\cite{siebenmorgen04} is featureless except for the presence of a
strong \NeII\ line and is very similar to the nucleus of M83.
%From this figure it is also evident that, in general, sources with PAH
%detection have a high 11.2\,PAH-to-continuum ratio while sources with
%no PAH detection have a very low 11.2\,PAH-to-continuum ratio as it
%was found in Fig.\,\ref{fig:dustvspah}.

%Note that those objects with
%high upper limits to the 11.2\,\micron\ PAH luminosity have very
%strong continua and consequently the integration time of their
%observations is low. Therefore, their spectra have low sensitivities,
%thus high upper limits but still low 11.2\,PAH-to-continuum ratios.

Starburst galaxies are very similar to Galactic \HII\ regions, thus we
have checked this relationship between the hardness of the radiation
field and the presence of PAHs using a large sample of Galactic \HII\
regions observed with ISO/SWS \citep{peeters:catalogue}. While all the
\HII\ regions in the catalogue show the presence of PAHs, only 4
of them have a \SIV/\NeII\ ratio $> 0.35$: W3A,
IRAS~12063$-$6259, IRAS~12073$-$6233 and IRAS~11143$-$6134.
%The hardness of the radiation field of these 4 sources is not related to
%the metallicity gradient in the Milky Way but rather to the presence
%of early O-stars.
Their ISO/SWS spectra clearly show PAH emission bands. However, we
have recently obtained TIMMI2 slit spectra for 2 of these sources,
IRAS~12063$-$6259 and IRAS~12073$-$6233. For neither of them we have
detected PAH emission bands in their long-slit spectra
(Mart\'{\i}n-Hern\'{a}ndez et al., in preparation), even though the
TIMMI2 slit did cross the ionisation front and hence the PDR where the PAH
emission originates from. As mentioned above, this suggests that the
PAH emission is probably diffuse and extended and hence not dominated
by the denser regions but rather by the exposed PDR and
ISM. This diffuse emission
might be difficult to detect by TIMMI2 due to its narrow slit.

The apparent relation between the hardness of the radiation field
and the intensity of the PAH emission is consistent with recent results by
\citet{Madden05}. These authors found a strong correlation between the
\NeIII/\NeII\ ratio, tracing the hardness of the radiation field, and
the PAH/dust intensity ratio for a wide range of objects, including
low-metallicity galaxies as well as Galactic \HII\ regions and other metal rich galaxies. They characterise the dust emission as the fitted, feature-free
continuum of their MIR spectra between 10 and 16~\micron. These authors
conclude that this hard
radiation field must play an important role in the destruction of PAHs in low
metallicity regions.
%However, PAHs originate inside the PDR, thus they are not co-located with the
%high energy photons found within \HII\ regions. Hence, there should
%not be a direct relation between the PAHs and the presence of these
%energetic photons traced by ratios such as \SIV/\NeII\ although
%one may expect a causual one.

The lack and/or weakness of the PAH emission bands in galaxies
or \HII\ regions with a hard radiation field (i.e. \SIV/\NeII\ $>
0.35$) may be due to a low PAH abundance as a consequence of e.g.
 (i) PAH dehydrogenation, i.e. the CH bond rupture due to the absorption of a UV photon (since we are using the 11.2\,\micron\ PAH
emission band as a tracer for the presence of PAHs);
 (ii) PAH destruction;
 (iii) the relative contribution of the different phases in the ISM or
 (iv) PAH-dust competition, i.e. PAHs are not being excited due to the
presence of dust located inside the \HII\ regions.
In principle, a low PAH abundance could also originate from a low PAH
formation rate due to a dust formation process that is different to
that occuring in our Galaxy as e.g. may be the case in low metallicity
galaxies.  However, we deem this possibility unlikely to explain the
general relation we find between hardness of the radiation field and
low PAH fluxes since our sample also includes Galactic \HII\ regions
without PAH detections.

Here we discuss the four possibilities mentioned above:

\paragraph{(i) PAH dehydrogenation:}
We have traced the presence or non-presence of PAHs in this work
through the detection of the 11.2\,\micron\ PAH band. To first order,
the PAH/FIR can be traced by the 11.2\,PAH/FIR since (1) dehydrogenation
of PAHs (which would effect the strength of this CH mode) does not
seem to be important (see below) and (2) the fraction of total PAH flux emitted
in the 11.2\,\micron\ PAH band varies only from 9 to 27 \% with an
average of 18$\pm$5\% \citep{Peeters02}. Indeed, space-based
observations suggest that in regions with low or no PAH emission (as
is the case of e.g. \iizw) the strength of {\it all} PAH emission
bands decreases simultaneously and not only that of the 11.2\,\micron\
PAH band \citep[e.g.][]{Madden00, Laurent00, Madden05}. In addition, ISO/SWS
observations of a large sample of sources as well as theoretical
calculations show that dehydrogenation of PAHs should have no effect
on the observed PAH spectrum \citep[for a summary, see][]{hony01}.

\paragraph{(ii) PAH destruction:}
In order to investigate PAH destruction as a consequence of the
hardness of the radiation field as traced by ratios such as
\SIV/\NeII, one should keep in mind that PAH emission originates
inside the PDR and thus is not co-located with the high energy photons
found within \HII\ regions.

The strength of the radiation field at
the location where the PAH emission originates, the PDR, is generally
expressed by the incident FUV flux between 6 and 13.6~eV measured in
units of the average interstellar radiation field \citep[1.6
10$^{-6}$\, W/m$^2$,][]{Habing:G0:68}. This quantity is called G$_0$.
We have estimated the values of G$_0$ for the objects in our
sample and those in the study by \cite{siebenmorgen04}.  They have
been calculated using the dust temperature, T$_{\rm d}$, derived from
ISO/LWS observations \citep{Negishi:01} and the relation between G$_0$
and T$_{\rm d}$ given by \citet{Hollenbach:91} with A$_{\rm V}$=0.5
\citep{Negishi:01}. In the cases where T$_{\rm d}$ is not known, we have
used the relation between the gas density, n, and the ratio
of the IRAS 60 and 100 \micron\ fluxes
%\citep[eq. 10 of][]{Negishi:01}
together with the relation between G$_0$ and n \citep{Negishi:01}. In
any case, the obtained G$_0$ is an average for the entire galaxy due
to the large LWS and IRAS apertures. Also note that these two methods
ignore metallicity effects and thus the values of G$_0$ we obtain are
less accurate in the case of low metallicity galaxies.

Log(G$_0$) is found to be in the range of 1.8--3.7~dex with an average
value of 3.0$\pm$0.4~dex. No correlation is found between G$_0$ and
the PAH strength or between G$_0$ and the presence or non-presence of
PAH emission bands in either the TIMMI2 or ISO
observations. Furthermore, this range in G$_0$ is similar to the
values typically seen in Galactic RNe \citep{YoungOwl:02}, whose IR
spectra are dominated by PAH emission bands. Hence, this suggests
that it is unlikely that the low 11.2\,PAH-to-continuum ratio reflects a
destruction of PAHs in these environments (i.e. PDRs).

We have to point out again that the G$_0$ values we estimate
correspond to the average G$_0$ of the PDRs in the galaxy because of
the large apertures (LWS, IRAS) we are considering and might not be
representative of the conditions in the small/unresolved sources we
are studying here. The above conclusion must be then taken with
caution.  For low metallicity galaxies it is found that the PDRs could
be restricted to very small dense clumps \citep{Madden00,
Galliano05}. If that is the case, the estimated G$_0$'s would then be
an average of such clumps.

\paragraph{(iii) Relative contribution of different regions :}
Rather than invoking a {\it special} PAH destruction, the low PAH
emission could just be a consequence of the relative contribution of
the different phases of the interstellar medium in the beam. For example, a
substantial pervasive, diffuse and highly ionised medium is found
throughout low metallicity galaxies \citep{Madden05} and - as known
from Galactic \HII\ regions \citep[e.g.][]{Tielens:anatomyorionbar:93,
Verstraete:m17:96} - PAHs do not survive inside highly ionised
mediums.  The presence of a diffuse ionised medium implies a smaller
contribution of the diffuse neutral interstellar medium (PDRs, ISM).
This neutral medium contributes considerably to the observed PAH
emission in starburst and normal galaxies \citep[see above and
e.g.][]{peeters04} and hence, such contribution to the PAH emission
would then be missing in low metallicity galaxies and would explain
the low PAH abundances in these galaxies.  This is consistent with
what we said earlier regarding the aperture effect on the detection of
PAHs: PAHs are more prominent in large aperture observations because
the relative contribution of the PDRs and ISM with respect to the
highly ionised medium of the \HII\ regions is larger.

\paragraph{(iv) PAH-dust competition:}
Another explanation might be found in the competition for FUV photons
by the dust grains and the PAH molecules. Dust grains are able to
reside inside the \HII\ region \citep[see e.g. M17, Fig. 3
of][]{Verstraete:m17:96}. When this occurs, the dust located inside
the \HII\ region absorbs a large fraction of photons before they reach
the PDR. In particular, the optical depth of dust in \HII\
regions, $\tau$(dust), is proportional to $n^{1/3}$. Hence, at high
densities, more of the FUV photons are absorbed by dust in the \HII\
region than at low densities. Moreover, a larger fraction of the
ionising photons is absorbed by the dust rather than the ionised
gas. As a result, PAHs, which are located outside of the
\HII\ regions, are less excited. In addition, the dust inside the
\HII\ region attains very high temperatures and consequently the
PAH-to-continuum ratio is lower. Indeed, Fig.~\ref{fig:dustvspah}
shows a relatively large spread in the 11.2\,PAH-to-continuum ratio
with the lowest ratios corresponding to the highly excited
sources. This is also nicely illustrated by the two different ISO
pointings towards the extended \HII\ region M17 plotted in
Fig.\,\ref{fig:dustvspah}, one inside the \HII\ region and the other
towards the molecular cloud. Thus, the absence of PAH emission in
highly excited sources might be due to a lower availability of
exciting photons due to the presence of dust within the \HII\
region.\\

Concluding, the cause of this relation between the hardness of the
radiation field and the presence or non-presence of PAHs {\it does not
necessarily imply PAH destruction}. Since PAHs originate inside
the PDR and thus are not co-located with the high energy photons found
within \HII\ regions, this relation might well be connected to the
presence of dust within the \HII\ region leading to a PAH-dust
competition for UV photons.
We consider also the scenery where the low PAH
emission could just be a consequence of the relative contribution of
the different phases of the interstellar medium, in particular, the presence
of a pervasive and highly ionised medium.

\subsection{The stellar content of the IR supernebula in \iizw}
\label{sect:cloudy}

It has been shown in Sect.~\ref{sect:iizw40} that practically most (if
not all) of the MIR line fluxes of \iizw\ measured by ISO is
confined to the compact (0\farcs5) source observed by TIMMI2. We can
use then the large number of MIR lines observed by ISO to constrain
properties of the ionising cluster following the analysis done with
NGC\,5253~C2 in our previous work \citep{martin:ngc5253}.

   \begin{figure*}[!ht]
   \centering \includegraphics[width=13.5cm]{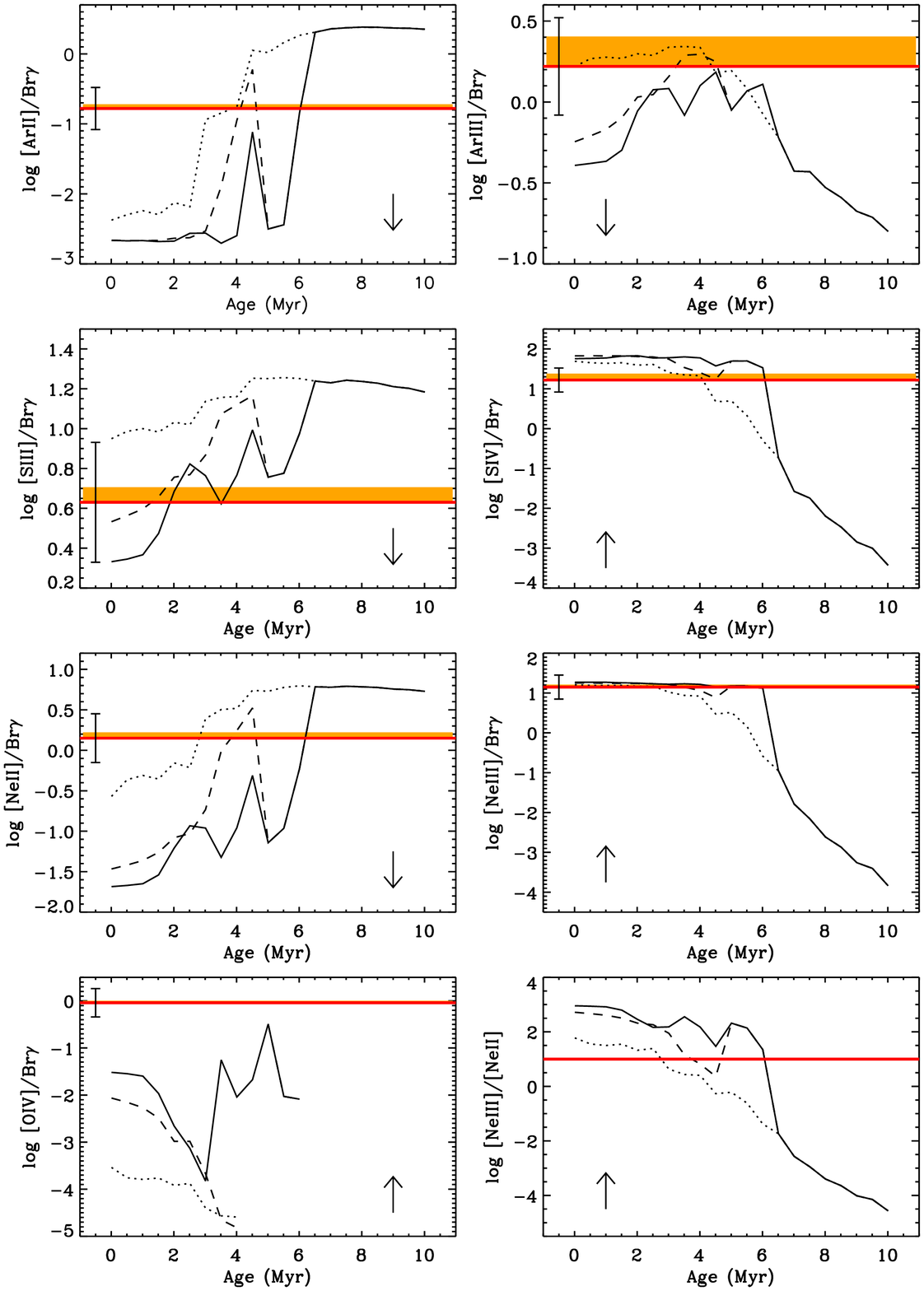}
%   \centering \includegraphics[width=10.7cm]{./plotdiagnost_1.eps}
   \caption{Variation of selected emission line ratios as a function
   of starburst age. The observed line ratios are indicated by an
   horizontal line.
   The \ArIII\ 9.0, \SIV\ 10.5 and \NeII\ 12.8~\micron\ lines have been
   measured with TIMMI2 (cf. Table~\ref{table:fluxes:iizw40}). The lines of
   \ArIII\ 7.0, \NeIII\ 15.5 and \SIII\ 18.7~\micron\ have been measured
   with ISO \citep{verma03}. Br$\gamma$ is being estimated from the
   2~cm free-free emission (see Sect.~\ref{sect:ab:iizw40})
   and we show as an error bar on the left side of each panel the effect
   of having under/overestimating this line flux by a factor of 2.
   Plots involving
   ratios with lines measured within the same aperture such as \SIV/\NeII\ and
   \SIV/\ArIII\ give the same result as that of \NeIII/\NeII\ vs age.
   The light band shows the effect of correcting the
   MIR lines for a visual extinction of 10 mag
   (cf. Sect~\ref{sect:iizw40:extinction}). The legend of the figure
   is the following: solid lines correspond to models with
   \mup=100\msun, dashed lines to models with \mup=50\msun\ and dotted
   lines to models with \mup=30\msun.  The nebular parameters of these
   models are $Q_0=10^{53}$~s$^{-1}$, $R_{\rm in}=4.3$~pc, $R_{\rm
   out}\sim 11$~pc, \den=1700~\cm\ and $\epsilon=1$, which correspond
   to $\log U\sim -0.05$. The effect of increasing the parameter U
   and/or decreasing the metallicity of the stellar cluster and/or
   adopting a steeper IMF and/or including internal dust is indicated
   by arrows.}
         \label{fig:cloudy}
   \end{figure*}

Stellar properties can be derived from nebular lines using
photoionisation models. These photoionisation models depend on the SED
of the ionising cluster (age and IMF), the local abundance and the ionisation
parameter, $U$. In particular, the parameter $U$, which depends on the
geometry, is usually difficult to constrain.

We compute sets of nebular models with the photoionisation code
CLOUDY\footnote{see http://thunder.pa.uky.edu/cloudy/}
\citep{ferland98} version 96.00 using MICE\footnote{see
http://isc.astro.cornell.edu/$\sim$spoon/mice.html}, the IDL interface
for CLOUDY created by H. Spoon. The computation is performed for a
static, spherically symmetric, ionisation bounded gas distribution
with an inner cavity.  We assume that the gas is uniformly distributed
in small clumps of constant density over the nebular volume and
occupies a fraction $\epsilon$ of the total volume.  The input
parameters of the photoionisation models are: the shape of the SED,
the ionising photon luminosity ($Q_0$), the electron
density ($n_{\rm e}$), the inner radius of the shell of ionised gas
($R_{\rm in}$), the filling factor ($\epsilon$), and the chemical
composition ($Z$).

We have used the evolutionary synthesis code Starburst99\footnote{see
http://www.stsci.edu/science/starburst99/} \citep{leitherer99} version
4.0 to model the integrated properties of the stellar cluster.
%This code is based on stellar evolution models of the Geneva group and uses
%enhanced mass loss tracks for masses above 12\,\msun\ \citep{meynet94}
%and standard mass loss tracks between 0.8 and 12\,\msun
%\citep{schaller92,schaerer93a,schaerer93b,charbonnel93}.  Starbust99
%follows the evolution in the H-R diagram of a stellar population whose
%composition is specified by a stellar initial mass function (IMF). At
%a given age, the integrated SED is obtained by summing over the
%contributions of all stars present and is built using: 1) the non-LTE
%WM-Basic stellar models \citep{pauldrach01, smith02} for O stars,
%which take into account the effects of stellar winds and line
%blanketing; 2) the fully line-blanketed models calculated with the
%CMFGEN code \citep{hillier98} for Wolf-Rayet stars; and 3) the
%plane-parallel LTE models by \cite{kurucz93} for the remaining stars
%that contribute to the continuum.
We assume an instantaneous burst of
star formation, a \cite{salpeter55} IMF with exponent $\alpha=2.35$
($dN/d\ln m \propto m^{1-\alpha}$), a lower mass cutoff $M_{\rm
low}=1$~M$_{\sun}$ and an upper mass cutoff $M_{\rm up}$ set to 30, 50
and 100~M$_{\sun}$. We also assume that the stars evolve from the main
sequence following the $Z=0.008$ stellar tracks. Stellar tracks with
$Z=0.006$, the metallicity measured for \iizw, are not
available but the effect of decreasing $Z$ will be discussed below.
We present models every 0.5 Myr for 10 Myr after the burst of star
formation.

The chemical composition of the gas is set to $Z=0.006$
(cf. Table~\ref{table:prop}).  We have adjusted the helium abundance
$Y$ according to $Y=Y_{\rm p}+(\Delta Y/\Delta Z)Z$, where $Y_{\rm
p}=0.24$ is the primordial helium abundance \citep{audouze87} and
$\Delta Y/\Delta Z=3$ is an observed constant \citep{pagel92}. To
arrive at the appropriate metal abundances, we have simply scaled the
solar values stored in the CLOUDY database.

A change of the rate of Lyman ionising photons, electron density, inner radius
of the shell and/or filling factor is equivalent to a change of the
ionisation parameter ($U$) defined by CLOUDY as:

\begin{equation}
U=Q_{0}/(4\pi R_{\rm in}^2 n_{\rm e}c)~.
\label{eq:u}
\end{equation}

There exists a relationship between $Q_0$, $n_{\rm e}$, the inner
radius $R_{\rm in}$, the outer radius $R_{\rm out}$ and the filling
factor $\epsilon$ given by \cite{martin:ngc5253}:

\begin{equation}
(1 - x^3) R_{\rm out}^3 \epsilon =
{ {3 Q_0} \over {4 \pi n_{\rm e}^2 \alpha_{\rm B}} }
\label{eq:x}
\end{equation}

\noindent
where $R_{\rm in}=xR_{\rm out}$ and $0 < x < 1$. Physical solutions of
this equation are those that satisfy the condition $\epsilon \le 1$.

Considering $Q_0=10^{53}$~s$^{-1}$, an electron density of $\sim
1700$~\cm\ and an outer radius $R_{\rm out}=11.15$~pc
(cf. Table.~\ref{table:prop}), together with the above condition, we
obtain that $R_{\rm in} \leq 0.38 \times R_{\rm out}=4.3$~pc, which
gives a value of $\log U \geq -0.05$ (cf. Eq.~\ref{eq:u}). More
compact geometries (i.e. with smaller $R_{\rm out}$ and bigger
densities) will have larger values of the ionisation parameter. We
will consider then a nebula with the following parameters:
$Q_0=10^{53}$~s$^{-1}$, $R_{\rm in}=4.3$~pc, \den$=1700$~\cm\ and
$\epsilon=1$. The effect of a larger ionisation factor will be
discussed.

Fig.~\ref{fig:cloudy} shows the results of the photoionisation models
as a function of time since the burst of star formation. {\it The same
conclusions as in the case of the super-star cluster C2 in NGC\,5253
can be reached}.  I.e. we find two possible solutions for the age and
upper mass cutoff of \iizw: 1) a young ($\la 3- 4$~ Myr) cluster with a low
upper mass cutoff \mup$< 50$\,\msun; and 2) a cluster of $\sim$ 4--7~Myr with
a standard high upper mass cutoff \mup$\sim 100$\,\msun.

%This result
%is confirmed by all the line ratio predictions shown in
%Fig.~\ref{fig:cloudy} except those of \ArIII/Br$\gamma$,
%\SIII/Br$\gamma$ and \OIV/Br$\gamma$. The models with \mup=100\,\msun\
%barely reproduce the observed \ArIII/Br$\gamma$, but they are only
%$\sim 0.1$~dex below the observations for an age of 5--6~Myr. In the
%case of \SIII/Br$\gamma$, the dependence of this ratio with age and
%upper mass cutoff is small ($\sim 0.8$~dex) and a small increase of
%the ionisation factor could reconcile this ratio with the others.

Basically, all line ratio predictions shown in Fig.~\ref{fig:cloudy}, except
\OIV/Br$\gamma$, are compatible with this result.
In the case of \ArIII/Br$\gamma$, \SIII/Br$\gamma$, and \NeII/Br$\gamma$,
these line ratios are not very sensitive
or their uncertainties are too large.

The \OIV\ line is known to be difficult to reproduce by pure stellar
photoionisation
\citep[][]{lutz98,schaerer99,martin:ngc5253},
but could also be related to non-stellar processes
\citep[c.f.][]{lutz98,viegas99},
as indicated  e.g.\ by recent IRS/Spitzer observations of NGC\,5253
by \citet{beirao06}. This underprediction of the \OIV\ line flux by the
models was also encountered in
the case of NGC\,5253~C2, where it was concluded that a modification
of the adopted SEDs close to and beyond the \HeII\ edge (note that the
ionisation edge for the creation of \ion{O}{iv}, 54.9~eV, is the
hightest of all the considered ions) was necessary in order to
reproduce all the MIR lines.

The increase of the ionisation parameter, whose effect in the
predictions is indicated by arrows in Fig.~\ref{fig:cloudy}, will only
accentuate the differences between the observations and the models
with \mup$=100$\,\msun. Consequently, if the nebula had a higher
parameter $U$, the upper mass cutoff of the IMF would need to be
revised downward in comparison with the models discussed here. The
same holds in the case of using SEDs with a lower metallicity or
adding internal dust in the nebula. These and other factors such as
varying the power law index of the IMF and the density law or
considering a matter bounded geometry for the nebula are discussed in
detailed in \cite{martin:ngc5253}. There it was shown that any of
these effects can reconcile the models with \mup$=100$\,\msun\ with
the observations.
An age $< 3-4$~Myr would agree with the optically thick thermal
bremsstrahlung origin of the radio emission associated with the IR knot
and the lack of supernova signatures
(cf. Sect.~\ref{sect:he2-10:general}). However, it would imply an IMF
with a low upper mass cutoff (\mup$\lesssim 50$\,\msun). The solution
of an older cluster of $\sim 4-7$~Myr would imply that it is possible
to contain such compact regions for a longer time that what it is
generally though while at the same time, hide the signatures of
supernovae.
The bright compact nebula in \iizw\ might thus be the second object, together
with NGC\,5253~C2, for which indications of a ``non-standard'' IMF
with a low upper mass cutoff \mup$\lesssim 50$\,\msun\ exist.
Whether this is a characteristic common in compact objects or
whether this is due to defects in the stellar and/or nebular modelling
remains to be clarified.

Generally fairly young ages are found from stellar populations studies
in the optical
\citep[e.g.][]{raimann00,kong03,westera04}.
In particular WR features indicate ages of $\sim$ 3--5 Myr
\citep[cf.][]{schaerer99,guseva00}.
However, it is not clear if and how this population relates to the one
responsible for the mid-IR line emission.

A solution that might lead to the lowering of the ionisation parameter and thus to the reconciliation of model and observations is the presence of multiple clusters within the compact core we are considering here. Indeed, the high spatial resolution observations at 2~cm presented by \cite{beck02} reveal 3 or 4 peaks within the central radio emission. They argue that these knots may simply be peaks of the more extended distribution, but also consider the possibility of young \HII\ regions with diameters of the order of $\sim 1$~pc. \cite{forster01} show that regions with multiple \HII\ regions are best modelled by a randomised distribution of clouds and clusters described by an effective ionisation parameter $U_{\rm eff}$. The computation of $U_{\rm eff}$ depends, however, on the detailed knowledge of the properties and distribution of the ionising stars and gas clouds. For the well known starburst galaxy M82, \cite{forster01} obtain a value of the effective
ionisation parameter lower than using the conventional definition (cf. Eq.~\ref{eq:u}) due to the important increase in surface area of the gas exposed to the Lyman continuum radiation field for the more realistic randomised distribution.
Indeed such a general tendency can be expected for multiple independent
\HII\ regions producing the same Lyman continuum flux as a single region.
However, the opposite trend may be the case in sufficiently compact environments
once the multiple \HII\ regions start to overlap, or if they provide a
non negligible external ionising radiation field (e.g. due to some
leakage). In this case the outer parts, host of emission from low ionisation
species, can be eroded or destroyed, leading hence to a global shift
toward emission of higher ionisation. Which situation applies to \iizw\
cannot be judged from the currently available observations.

\section{Conclusions}
\label{sect:conclusions}

%{\bf DANIEL $\rightarrow$ From Xander: does it make sense to compare 5253C2 results to II %zwicky 40 in
%some detail and perhaps draw some comparison to early universe star
%formation? Could we do this?}

We have presented $N$-band spectra (8--13~\micron) of
some locations in three starburst galaxies. In particular, the two
galactic nuclei of the spiral galaxy NGC\,3256 (North and South), the
compact infrared supernebula in the dwarf galaxy \iizw\ and the two
brightest infrared knots in the central starburst of the WR galaxy
\henize\ (named A and C). These spectra have been obtained with TIMMI2
on the ESO 3.6\,m telescope.

The spectra show an ample variety in terms of continuum,
lines and molecular band strength. The two nuclei of NGC\,3256 and the
two IR knots in \henize\ show a rising dust continuum, PAH bands
and a strong \NeII\ line. On the contrary, the infrared knot in \iizw\
is characterised by a rather flat continuum, a strong \SIV\ line and
do not show the presence of PAH bands.

We demonstrated the great value of these type of data at constraining
properties such as the extinction in the MIR and
metallicity. Results to highlight at this respect are the following.
(1) There is an indication of supersolar metallicity
for the two nuclei of NGC\,3256, while a possible enhancement of neon with
respect to oxygen is found
in the two IR knots observed in \henize.
However, if the latest upward revision of the solar Ne abundance
by $\sim$0.4--0.5~dex suggested from helioseismology
and from solar type stars is confirmed,
our Ne abundance estimates could well be compatible with solar or even
(slightly) sub-solar.
(2) Around 60\% of the sulphur in the bright IR
knot in \iizw\ could be in the form of S$^{4+}$, indicating the presence
of photons in excess of 47~eV.
We note however that these abundance estimates largely depend on the
assumed flux for the Br$\gamma$ line. Hence, they must be
taken with certain caution since the MIR and Br$\gamma$ line fluxes
are not measured on the same aperture.

We have shown the importance of high-spatial observations of
extra-galactic sources when compared to observations obtained with
larger apertures such as ISO. Only such observations provide accurate
measurements of the line and molecular band fluxes emitted by the
cluster or galactic nucleus.

We adapted the MIR/FIR diagnostics of \citet{peeters04} in order to
use the 11.2\,\micron\ PAH band, accessible to ground-based
observations, instead of the 6.2\,\micron\ one and
compared our observations and those of \citet{siebenmorgen04} with
Galactic \HII\ Regions.  We find that the extra-galactic nuclei and
star clusters observed at high spatial resolution (as is the case of
the TIMMI2 observations) are closer in this diagram to compact
\HII\ regions, while galaxies observed by large apertures
such as ISO are nearer to exposed PDRs such as
Orion. This is certainly due to the aperture
difference, where the much larger ISO aperture likely includes much of the
surrounding PDRs while the
TIMMI2 slit measures mainly the central emission of the \HII\ region.

We have also found a dependence between the PAH presence or
non-presence and the hardness of the radiation field, as measured by
the \SIV/\NeII\ ratio: sources with PAH emission have a \SIV/\NeII\
ratio $\lesssim$ 0.35, while sources which do not show PAHs have line
ratios above 0.35. We investigated possible origins for this relation
and conclude that it does not necessarily
imply PAH destruction, but could also be explained by the PAH-dust
competition for FUV photons.  Dust grains probably absorb a large
fraction of the UV photons before they reach the PDR where the PAHs
reside, what likely cause a lower excitation of the PAHs.
We have also considered the scenario where the low PAH
emission could just be a consequence of the relative contribution of
the different phases of the interstellar medium, in particular, the presence
of a pervasive and highly ionised medium.

Finally, these data proved useful at constraining properties
(age, IMF, etc.) of the stellar content. Following the analysis
previously done with the IR supernebulae C2 in NGC\,5253
\citep{martin:ngc5253}, we have constrained the stellar content of the
IR compact knot in \iizw\ using the MIR fine-structure lines and
strong restrictions on the nebular geometry. The same conclusions as
in the case of the super-star cluster C2 can be reached.  I.e. we find
two possible solutions for the age and upper mass cutoff of \iizw: 1)
a young ($\la 3- 4$~ Myr) cluster with a low upper mass cutoff \mup$<
50$\,\msun; and 2) a cluster of 4--7~Myr with a standard high upper
mass cutoff \mup$\sim 100$\,\msun.
We show however that the presence of multiple clusters within the compact 
IR core
we are considering here, suggested by radio high spatial resolution 
observations,
might lead to a lowering of the ionisation parameter and thus to a
reconciliation between the observations and the models with a standard 
upper mass
cutoff of 100\,\msun.

\begin{acknowledgements}
We first thank the referee for his/her critical reading and
highly constructive comments.
DS wishes to thank Bernhard Brandl, Paul Crowther, Kelsey Johnson,
Christophe Morisset, and Bill Vacca for interesting discussions.
NLMH thanks the support given by the ``Juan de la Cierva'' programme of
the Spanish Ministry of Education and Science.  
EP acknowledges the support of the The National Research Council.
Part of this work was funded by the Swiss National Science Foundation.
\end{acknowledgements}

%\bibliographystyle{aa}
%\bibliography{/scratch/leticia/Bibliography/biblio,/scratch/leticia/Bibliography/els}

\begin{thebibliography}{121}
\expandafter\ifx\csname natexlab\endcsname\relax\def\natexlab#1{#1}\fi

\bibitem[{{Aguero} \& {Lipari}(1991)}]{aguero91}
{Aguero}, E.~L. \& {Lipari}, S.~L. 1991, Ap\&SS, 175, 253

\bibitem[{{Allamandola} {et~al.}(1989){Allamandola}, {Tielens}, \&
  {Barker}}]{allamandola89}
{Allamandola}, L.~J., {Tielens}, A.~G.~G.~M., \& {Barker}, J.~R. 1989, ApJS,
  71, 733

\bibitem[{{Allen} {et~al.}(1976){Allen}, {Wright}, \& {Goss}}]{allen76}
{Allen}, D.~A., {Wright}, A.~E., \& {Goss}, W.~M. 1976, MNRAS, 177, 91

\bibitem[{{Allende Prieto} {et~al.}(2001){Allende Prieto}, {Lambert}, \&
  {Asplund}}]{allende01}
{Allende Prieto}, C., {Lambert}, D.~L., \& {Asplund}, M. 2001, ApJ Lett., 556,
  L63

\bibitem[{{Alonso-Herrero} {et~al.}(2004){Alonso-Herrero}, {Takagi}, {Baker},
  {Rieke}, {Rieke}, {Imanishi}, \& {Scoville}}]{alonso04}
{Alonso-Herrero}, A., {Takagi}, T., {Baker}, A.~J., {et~al.} 2004, ApJ, 612,
  222

\bibitem[{{Antia} \& {Basu}(2005)}]{antia05}
{Antia}, H.~M. \& {Basu}, S. 2005, ApJ Lett., 620, L129

\bibitem[{{Audouze}(1987)}]{audouze87}
{Audouze}, J. 1987, in IAU Symp. 124: Observational Cosmology, Vol. 124,
  89--115

\bibitem[{{Bahcall} {et~al.}(2005){Bahcall}, {Basu}, \& M.}]{bahcall05}
{Bahcall}, J.~N., {Basu}, S., \& M., S.~A. 2005, {accepted for publication in
  ApJ, see astro-ph/0502563}

\bibitem[{{Baldwin} {et~al.}(1982){Baldwin}, {Spinrad}, \&
  {Terlevich}}]{baldwin82}
{Baldwin}, J.~A., {Spinrad}, H., \& {Terlevich}, R. 1982, MNRAS, 198, 535

\bibitem[{{Beck} {et~al.}(1997){Beck}, {Kelly}, \& {Lacy}}]{beck97}
{Beck}, S.~C., {Kelly}, D.~M., \& {Lacy}, J.~H. 1997, AJ, 114, 585

\bibitem[{{Beck} {et~al.}(2001){Beck}, {Turner}, \& {Gorjian}}]{beck01}
{Beck}, S.~C., {Turner}, J.~L., \& {Gorjian}, V. 2001, AJ, 122, 1365

\bibitem[{{Beck} {et~al.}(2002){Beck}, {Turner}, {Langland-Shula}, {Meier},
  {Crosthwaite}, \& {Gorjian}}]{beck02}
{Beck}, S.~C., {Turner}, J.~L., {Langland-Shula}, L.~E., {et~al.} 2002, AJ,
  124, 2516

\bibitem[{{Beirao} {et~al.}(2006){Beirao}, {Brandl}, {Devost}, {Smith}, {Hao},
  \& {Houck}}]{beirao06}
{Beirao}, P., {Brandl}, B.~R., {Devost}, D., {et~al.} 2006, {submitted to ApJL}

\bibitem[{{B\"oker} {et~al.}(1997){B\"oker}, {Storey}, {Krabbe}, \&
  {Lehmann}}]{boker97}
{B\"oker}, T., {Storey}, J.~W.~V., {Krabbe}, A., \& {Lehmann}, T. 1997, PASP,
  109, 827

\bibitem[{{Cabanac} {et~al.}(2005){Cabanac}, {Vanzi}, \& {Sauvage}}]{cabanac05}
{Cabanac}, R.~A., {Vanzi}, L., \& {Sauvage}, M. 2005, ApJ, 631, 252

\bibitem[{{Clavel} {et~al.}(2000){Clavel}, {Schulz}, {Altieri}, {Barr},
  {Claes}, {Heras}, {Leech}, {Metcalfe}, \& {Salama}}]{clavel00}
{Clavel}, J., {Schulz}, B., {Altieri}, B., {et~al.} 2000, A\&A, 357, 839

\bibitem[{{Cohen} {et~al.}(1999){Cohen}, {Walker}, {Carter}, {Hammersley},
  {Kidger}, \& {Noguchi}}]{cohen99}
{Cohen}, M., {Walker}, R.~G., {Carter}, B., {et~al.} 1999, ApJ, 117, 1864

\bibitem[{{Conti}(1991)}]{conti91}
{Conti}, P.~S. 1991, ApJ, 377, 115

\bibitem[{{Conti} \& {Vacca}(1994)}]{conti94}
{Conti}, P.~S. \& {Vacca}, W.~D. 1994, ApJ Lett., 423, L97

\bibitem[{{Coziol} {et~al.}(2001){Coziol}, {Doyon}, \& {Demers}}]{coziol01}
{Coziol}, R., {Doyon}, R., \& {Demers}, S. 2001, MNRAS, 325, 1081

\bibitem[{{Davies} {et~al.}(1998){Davies}, {Sugai}, \& {Ward}}]{davies98}
{Davies}, R.~I., {Sugai}, H., \& {Ward}, M.~J. 1998, MNRAS, 295, 43

\bibitem[{{de Vaucouleurs} \& {de Vaucouleurs}(1961)}]{deVaucouleurs61}
{de Vaucouleurs}, G. \& {de Vaucouleurs}, A. 1961, Mem. R. Astron. Soc., 68, 69

\bibitem[{{Deeg} {et~al.}(1993){Deeg}, {Brinks}, {Duric}, {Klein}, \&
  {Skillman}}]{deeg93}
{Deeg}, H., {Brinks}, E., {Duric}, N., {Klein}, U., \& {Skillman}, E. 1993,
  ApJ, 410, 626

\bibitem[{{Doyon} {et~al.}(1994){Doyon}, {Joseph}, \& {Wright}}]{doyon94}
{Doyon}, R., {Joseph}, R.~D., \& {Wright}, G.~S. 1994, ApJ, 421, 101

\bibitem[{{Doyon} {et~al.}(1992){Doyon}, {Puxley}, \& {Joseph}}]{doyon92}
{Doyon}, R., {Puxley}, P.~J., \& {Joseph}, R.~D. 1992, ApJ, 397, 117

\bibitem[{{Draine}(1985)}]{draine85}
{Draine}, B.~T. 1985, ApJS, 57, 587

\bibitem[{{Draine} \& {Lee}(1984)}]{draine84}
{Draine}, B.~T. \& {Lee}, H.~M. 1984, ApJ, 285, 89

\bibitem[{{Drake} \& {Testa}(2005)}]{drake05}
{Drake}, J.~J. \& {Testa}, P. 2005, Nature, 436, 525

\bibitem[{{F{\" o}rster Schreiber} {et~al.}(2001){F{\" o}rster Schreiber},
  {Genzel}, {Lutz}, {Kunze}, \& {Sternberg}}]{forster01}
{F{\" o}rster Schreiber}, N.~M., {Genzel}, R., {Lutz}, D., {Kunze}, D., \&
  {Sternberg}, A. 2001, ApJ, 552, 544

\bibitem[{{Ferland} {et~al.}(1998){Ferland}, {Korista}, {Verner}, {Ferguson},
  {Kingdon}, \& {Verner}}]{ferland98}
{Ferland}, G.~J., {Korista}, K.~T., {Verner}, D.~A., {et~al.} 1998, PASP, 110,
  761

\bibitem[{{Forbes} \& {Ward}(1993)}]{forbes93}
{Forbes}, D.~A. \& {Ward}, M.~J. 1993, ApJ, 416, 150

\bibitem[{{Galliano} {et~al.}(2005){Galliano}, {Madden}, {Jones}, {Wilson}, \&
  {Bernard}}]{Galliano05}
{Galliano}, F., {Madden}, S.~C., {Jones}, A.~P., {Wilson}, C.~D., \& {Bernard},
  J.-P. 2005, \aap, 434, 867

\bibitem[{{Genzel} \& {Cesarsky}(2000)}]{genzel00}
{Genzel}, R. \& {Cesarsky}, C.~J. 2000, ARA\&A, 38, 761

\bibitem[{{Genzel} {et~al.}(1998){Genzel}, {Lutz}, {Sturm}, {Egami}, {Kunze},
  {Moorwood}, {Rigopoulou}, {Spoon}, {Sternberg}, {Tacconi-Garman}, {Tacconi},
  \& {Thatte}}]{genzel98}
{Genzel}, R., {Lutz}, D., {Sturm}, E., {et~al.} 1998, ApJ, 498, 579

\bibitem[{{Glass} \& {Moorwood}(1985)}]{glass85}
{Glass}, I.~S. \& {Moorwood}, A.~F.~M. 1985, MNRAS, 214, 429

\bibitem[{{Gorjian} {et~al.}(2001){Gorjian}, {Turner}, \& {Beck}}]{gorjian01}
{Gorjian}, V., {Turner}, J.~L., \& {Beck}, S.~C. 2001, ApJ Lett., 554, L29

\bibitem[{{Graham} {et~al.}(1984){Graham}, {Wright}, {Meikle}, {Joseph}, \&
  {Bode}}]{graham84}
{Graham}, J.~R., {Wright}, G.~S., {Meikle}, W.~P.~S., {Joseph}, R.~D., \&
  {Bode}, M.~F. 1984, Nature, 310, 213

\bibitem[{{Guseva} {et~al.}(2000){Guseva}, {Izotov}, \& {Thuan}}]{guseva00}
{Guseva}, N.~G., {Izotov}, Y.~I., \& {Thuan}, T.~X. 2000, ApJ, 531, 776

\bibitem[{{Habing}(1968)}]{Habing:G0:68}
{Habing}, H.~J. 1968, "Bulletin of the Astronomical Institute of the
  Netherlands", 19, 421

\bibitem[{{Ho} {et~al.}(1990){Ho}, {Beck}, \& {Turner}}]{ho90}
{Ho}, P.~T.~P., {Beck}, S.~C., \& {Turner}, J.~L. 1990, ApJ, 349, 57

\bibitem[{{Hollenbach} {et~al.}(1991){Hollenbach}, {Takahashi}, \&
  {Tielens}}]{Hollenbach:91}
{Hollenbach}, D.~J., {Takahashi}, T., \& {Tielens}, A.~G.~G.~M. 1991, ApJ, 377,
  192

\bibitem[{{Hony} {et~al.}(2001){Hony}, {Van Kerckhoven}, {Peeters}, {Tielens},
  {Hudgins}, \& {Allamandola}}]{hony01}
{Hony}, S., {Van Kerckhoven}, C., {Peeters}, E., {et~al.} 2001, A\&A, 370, 1030

\bibitem[{{Horne}(1986)}]{horne86}
{Horne}, K. 1986, PASP, 98, 609

\bibitem[{{Hunt} {et~al.}(2005){Hunt}, {Bianchi}, \& {Maiolino}}]{hunt05}
{Hunt}, L., {Bianchi}, S., \& {Maiolino}, R. 2005, A\&A, 434, 849

\bibitem[{{Jaffe} {et~al.}(1978){Jaffe}, {Perola}, \& {Tarenghi}}]{jaffe78}
{Jaffe}, W.~J., {Perola}, G.~C., \& {Tarenghi}, M. 1978, ApJ, 224, 808

\bibitem[{{Johansson}(1987)}]{johansson87}
{Johansson}, I. 1987, A\&A, 182, 179

\bibitem[{{Johnson}(2005)}]{johnson05}
{Johnson}, K.~E. 2005, in IAU Symposium 227, ed. R.~{Cesaroni}, M.~{Felli},
  E.~{Churchwell}, \& M.~{Walmsley}, 413--422

\bibitem[{{Johnson} \& {Kobulnicky}(2003)}]{johnson03}
{Johnson}, K.~E. \& {Kobulnicky}, H.~A. 2003, ApJ, 597, 923

\bibitem[{{Joy} \& {Lester}(1988)}]{joy88}
{Joy}, M. \& {Lester}, D.~F. 1988, ApJ, 331, 145

\bibitem[{{Kawara} {et~al.}(1989){Kawara}, {Nishida}, \& {Phillips}}]{kawara89}
{Kawara}, K., {Nishida}, M., \& {Phillips}, M.~M. 1989, ApJ, 337, 230

\bibitem[{{Klein} {et~al.}(1991){Klein}, {Weiland}, \& {Brinks}}]{klein91}
{Klein}, U., {Weiland}, H., \& {Brinks}, E. 1991, A\&A, 246, 323

\bibitem[{{Kobulnicky} {et~al.}(1995){Kobulnicky}, {Dickey}, {Sargent}, {Hogg},
  \& {Conti}}]{kobulnicky95}
{Kobulnicky}, H.~A., {Dickey}, J.~M., {Sargent}, A.~I., {Hogg}, D.~E., \&
  {Conti}, P.~S. 1995, AJ, 110, 116

\bibitem[{{Kobulnicky} \& {Johnson}(1999)}]{kobulnicky99}
{Kobulnicky}, H.~A. \& {Johnson}, K.~E. 1999, ApJ, 527, 154

\bibitem[{{Kong} {et~al.}(2003){Kong}, {Charlot}, {Weiss}, \& {Cheng}}]{kong03}
{Kong}, X., {Charlot}, S., {Weiss}, A., \& {Cheng}, F.~Z. 2003, A\&A, 403, 877

\bibitem[{{Kotilainen} {et~al.}(1996){Kotilainen}, {Moorwood}, {Ward}, \&
  {Forbes}}]{kotilainen96}
{Kotilainen}, J.~K., {Moorwood}, A.~F.~M., {Ward}, M.~J., \& {Forbes}, D.~A.
  1996, A\&A, 305, 107

\bibitem[{{Kurtz} {et~al.}(1994){Kurtz}, {Churchwell}, \& {Wood}}]{kurtz94}
{Kurtz}, S., {Churchwell}, E., \& {Wood}, D.~O.~S. 1994, ApJS, 91, 659

\bibitem[{{Laurent} {et~al.}(2000){Laurent}, {Mirabel}, {Charmandaris},
  {Gallais}, {Madden}, {Sauvage}, {Vigroux}, \& {Cesarsky}}]{Laurent00}
{Laurent}, O., {Mirabel}, I.~F., {Charmandaris}, V., {et~al.} 2000, A\&A, 359,
  887

\bibitem[{{Leitherer} {et~al.}(1999){Leitherer}, {Schaerer}, {Goldader},
  {Delgado}, {Robert}, {Kune}, {de Mello}, {Devost}, \&
  {Heckman}}]{leitherer99}
{Leitherer}, C., {Schaerer}, D., {Goldader}, J.~D., {et~al.} 1999, ApJS, 123, 3

\bibitem[{{L{\'{\i}}pari} {et~al.}(2000){L{\'{\i}}pari}, {D{\'{\i}}az},
  {Taniguchi}, {Terlevich}, {Dottori}, \& {Carranza}}]{lipari00}
{L{\'{\i}}pari}, S., {D{\'{\i}}az}, R., {Taniguchi}, Y., {et~al.} 2000, AJ,
  120, 645

\bibitem[{{L{\'{\i}}pari} {et~al.}(2004){L{\'{\i}}pari}, {D{\'{\i}}az},
  {Forte}, {Terlevich}, {Taniguchi}, {Aguero}, {Alonso-Herrero}, {Mediavilla},
  \& {Zepf}}]{lipari04}
{L{\'{\i}}pari}, S.~L., {D{\'{\i}}az}, R.~J., {Forte}, J.~C., {et~al.} 2004,
  MNRAS, 354, L1

\bibitem[{{Lira} {et~al.}(2002){Lira}, {Ward}, {Zezas}, {Alonso-Herrero}, \&
  {Ueno}}]{lira02}
{Lira}, P., {Ward}, M., {Zezas}, A., {Alonso-Herrero}, A., \& {Ueno}, S. 2002,
  MNRAS, 330, 259

\bibitem[{{Lutz} {et~al.}(1998){Lutz}, {Kunze}, {Spoon}, \&
  {Thornley}}]{lutz98}
{Lutz}, D., {Kunze}, D., {Spoon}, H.~W.~W., \& {Thornley}, M.~D. 1998, A\&A,
  333, L75

\bibitem[{{Madden}(2000)}]{Madden00}
{Madden}, S.~C. 2000, New Astronomy Review, 44, 249

\bibitem[{{Madden} {et~al.}(2005){Madden}, {Galliano}, Jones, \&
  {Sauvage}}]{Madden05}
{Madden}, S.~C., {Galliano}, F., Jones, A., \& {Sauvage}, M. 2005, {to appear
  in A\&A}

\bibitem[{{Mart{\'{\i}}n-Hern{\' a}ndez} {et~al.}(2003){Mart{\'{\i}}n-Hern{\'
  a}ndez}, {van der Hulst}, \& {Tielens}}]{martin:atca:gal}
{Mart{\'{\i}}n-Hern{\' a}ndez}, N.~L., {van der Hulst}, J.~M., \& {Tielens},
  A.~G.~G.~M. 2003, A\&A, 407, 957

\bibitem[{{Mart\'{\i}n-Hern\'{a}ndez}
  {et~al.}(2002){Mart\'{\i}n-Hern\'{a}ndez}, {Peeters}, {Morisset}, {Tielens},
  {Cox}, {Roelfsema}, {Baluteau}, {Schaerer}, {Mathis}, {Damour}, {Churhwell},
  \& {Kessler}}]{martin:paperii}
{Mart\'{\i}n-Hern\'{a}ndez}, N.~L., {Peeters}, E., {Morisset}, C., {et~al.}
  2002, A\&A, 381, 606

\bibitem[{{Mart{\'{\i}}n-Hern{\'a}ndez}
  {et~al.}(2005){Mart{\'{\i}}n-Hern{\'a}ndez}, {Schaerer}, \&
  {Sauvage}}]{martin:ngc5253}
{Mart{\'{\i}}n-Hern{\'a}ndez}, N.~L., {Schaerer}, D., \& {Sauvage}, M. 2005,
  A\&A, 429, 449

\bibitem[{{Masegosa} {et~al.}(1994){Masegosa}, {Moles}, \&
  {Campos-Aguilar}}]{masegosa94}
{Masegosa}, J., {Moles}, M., \& {Campos-Aguilar}, A. 1994, ApJ, 420, 576

\bibitem[{{Mathis}(1990)}]{mathis90}
{Mathis}, J.~S. 1990, ARA\&A, 28, 37

\bibitem[{{Moorwood}(1986)}]{moorwood86}
{Moorwood}, A.~F.~M. 1986, A\&A, 166, 4

\bibitem[{{Moorwood} \& {Oliva}(1994)}]{moorwood94}
{Moorwood}, A.~F.~M. \& {Oliva}, E. 1994, ApJ, 429, 602

\bibitem[{{Moran} {et~al.}(1999){Moran}, {Lehnert}, \& {Helfand}}]{moran99}
{Moran}, E.~C., {Lehnert}, M.~D., \& {Helfand}, D.~J. 1999, ApJ, 526, 649

\bibitem[{{Neff} {et~al.}(2003){Neff}, {Ulvestad}, \& {Campion}}]{neff03}
{Neff}, S.~G., {Ulvestad}, J.~S., \& {Campion}, S.~D. 2003, ApJ, 599, 1043

\bibitem[{{Negishi} {et~al.}(2001){Negishi}, {Onaka}, {Chan}, \&
  {Roellig}}]{Negishi:01}
{Negishi}, T., {Onaka}, T., {Chan}, K.-W., \& {Roellig}, T.~L. 2001, A\&A, 375,
  566

\bibitem[{{Norris} \& {Forbes}(1995)}]{norris95}
{Norris}, R.~P. \& {Forbes}, D.~A. 1995, ApJ, 446, 594

\bibitem[{{Pagel}(1992)}]{pagel92}
{Pagel}, B.~E.~J. 1992, in IAU Symp. 149: The Stellar Populations of Galaxies,
  Vol. 149, 133

\bibitem[{{Pagel} {et~al.}(1992){Pagel}, {Simonson}, {Terlevich}, \&
  {Edmunds}}]{pagel92b}
{Pagel}, B.~E.~J., {Simonson}, E.~A., {Terlevich}, R.~J., \& {Edmunds}, M.~G.
  1992, MNRAS, 255, 325

\bibitem[{{Peeters} {et~al.}(2004{\natexlab{a}}){Peeters}, {Allamandola},
  {Hudgins}, {Hony}, \& {Tielens}}]{peeters04:review}
{Peeters}, E., {Allamandola}, L.~J., {Hudgins}, D.~M., {Hony}, S., \&
  {Tielens}, A.~G.~G.~M. 2004{\natexlab{a}}, in Astrophysics of Dust, ed. A.~N.
  {Witt}, G.~C. {Clayton}, \& B.~T. {Draine}, Astronomical Society of the
  Pacific, vol 309, 141

\bibitem[{{Peeters} {et~al.}(2002{\natexlab{a}}){Peeters}, {Hony}, {Van
  Kerckhoven}, {Tielens}, {Allamandola}, {Hudgins}, \&
  {Bauschlicher}}]{Peeters02}
{Peeters}, E., {Hony}, S., {Van Kerckhoven}, C., {et~al.} 2002{\natexlab{a}},
  A\&A, 390, 1089

\bibitem[{{Peeters} {et~al.}(2002{\natexlab{b}}){Peeters},
  {Mart\'{\i}n-Hern\'{a}ndez}, {Damour}, {Cox}, {Roelfsema}, {Baluteau},
  {Tielens}, {Churchwell}, {Kessler}, {Mathis}, {Morisset}, \&
  {Schaerer}}]{peeters:catalogue}
{Peeters}, E., {Mart\'{\i}n-Hern\'{a}ndez}, N.~L., {Damour}, F., {et~al.}
  2002{\natexlab{b}}, A\&A, 381, 571

\bibitem[{{Peeters} {et~al.}(2004{\natexlab{b}}){Peeters}, {Spoon}, \&
  {Tielens}}]{peeters04}
{Peeters}, E., {Spoon}, H.~W.~W., \& {Tielens}, A.~G.~G.~M. 2004{\natexlab{b}},
  ApJ, 613, 986

\bibitem[{{P\'erez-Montero} \& {D{\'{\i}}az}(2003)}]{perez03}
{P\'erez-Montero}, E. \& {D{\'{\i}}az}, A.~I. 2003, MNRAS, 346, 105

\bibitem[{{Phillips} {et~al.}(1984){Phillips}, {Aitken}, \&
  {Roche}}]{phillips84}
{Phillips}, M.~M., {Aitken}, D.~K., \& {Roche}, P.~F. 1984, MNRAS, 207, 25

\bibitem[{{Pilyugin}(2001)}]{pilyugin01}
{Pilyugin}, L.~S. 2001, A\&A, 369, 594

\bibitem[{{Puget} \& {L\'eger}(1989)}]{puget89}
{Puget}, J.~L. \& {L\'eger}, A. 1989, ARA\&A, 27, 161

\bibitem[{{Raimann} {et~al.}(2000){Raimann}, {Bica}, {Storchi-Bergmann},
  {Melnick}, \& {Schmitt}}]{raimann00}
{Raimann}, D., {Bica}, E., {Storchi-Bergmann}, T., {Melnick}, J., \& {Schmitt},
  H. 2000, MNRAS, 314, 295

\bibitem[{{Rieke} \& {Lebofsky}(1985)}]{rieke85}
{Rieke}, G.~H. \& {Lebofsky}, M.~J. 1985, ApJ, 288, 618

\bibitem[{{Rieke} \& {Low}(1972)}]{rieke72}
{Rieke}, G.~H. \& {Low}, F.~J. 1972, ApJ Lett., 176, L95

\bibitem[{{Rigby} \& {Rieke}(2004)}]{rigby04}
{Rigby}, J.~R. \& {Rieke}, G.~H. 2004, ApJ, 606, 237

\bibitem[{{Rigopoulou} {et~al.}(1996){Rigopoulou}, {Lutz}, {Genzel}, {Egami},
  {Kunze}, {Sturm}, {Feuchtgruber}, {Schaeidt}, {Bauer}, {Sternberg}, {Netzer},
  {Moorwood}, \& {de Graauw}}]{rigopoulou96}
{Rigopoulou}, D., {Lutz}, D., {Genzel}, R., {et~al.} 1996, A\&A, 315, L125

\bibitem[{{Roche} \& {Aitken}(1984)}]{roche84}
{Roche}, P.~F. \& {Aitken}, D.~K. 1984, MNRAS, 208, 481

\bibitem[{{Roche} {et~al.}(1991){Roche}, {Aitken}, {Smith}, \&
  {Ward}}]{roche91}
{Roche}, P.~F., {Aitken}, D.~K., {Smith}, C.~H., \& {Ward}, M.~J. 1991, MNRAS,
  248, 606

\bibitem[{{Rowan-Robinson} \& {Crawford}(1989)}]{rowan-robinson89}
{Rowan-Robinson}, M. \& {Crawford}, J. 1989, MNRAS, 238, 523

\bibitem[{{Rubin} {et~al.}(1988){Rubin}, {Simpson}, {Erickson}, \&
  {Haas}}]{rubin88}
{Rubin}, R.~H., {Simpson}, J.~P., {Erickson}, E.~F., \& {Haas}, M.~R. 1988,
  ApJ, 327, 377

\bibitem[{{Salpeter}(1955)}]{salpeter55}
{Salpeter}, E.~E. 1955, ApJ, 121, 161

\bibitem[{{Sargent} \& {Searle}(1970)}]{sargent70}
{Sargent}, W.~L.~W. \& {Searle}, L. 1970, ApJ Lett., 162, L155

\bibitem[{{Sauvage} {et~al.}(1997){Sauvage}, {Thuan}, \& {Lagage}}]{sauvage97}
{Sauvage}, M., {Thuan}, T.~X., \& {Lagage}, P.~O. 1997, A\&A, 325, 98

\bibitem[{{Schaerer} \& {Stasi{\' n}ska}(1999)}]{schaerer99}
{Schaerer}, D. \& {Stasi{\' n}ska}, G. 1999, A\&A, 345, L17

\bibitem[{{Sellgren} {et~al.}(1990){Sellgren}, {Tokunaga}, \&
  {Nakada}}]{Sellgren:90}
{Sellgren}, K., {Tokunaga}, A.~T., \& {Nakada}, Y. 1990, ApJ, 349, 120

\bibitem[{{Siebenmorgen} {et~al.}(2004){Siebenmorgen}, {Kr{\" u}gel}, \&
  {Spoon}}]{siebenmorgen04}
{Siebenmorgen}, R., {Kr{\" u}gel}, E., \& {Spoon}, H.~W.~W. 2004, A\&A, 414,
  123

\bibitem[{{Sramek} \& {Weedman}(1986)}]{sramek86}
{Sramek}, R.~A. \& {Weedman}, D.~W. 1986, ApJ, 302, 640

\bibitem[{{Stasi{\'n}ska}(2005)}]{stasinska05}
{Stasi{\'n}ska}, G. 2005, \aap, 434, 507

\bibitem[{{Storchi-Bergmann} {et~al.}(1995){Storchi-Bergmann}, {Kinney}, \&
  {Challis}}]{storchi95}
{Storchi-Bergmann}, T., {Kinney}, A.~L., \& {Challis}, P. 1995, ApJS, 98, 103

\bibitem[{{Storey} \& {Hummer}(1995)}]{storey95}
{Storey}, P.~J. \& {Hummer}, D.~G. 1995, MNRAS, 272, 41

\bibitem[{{Thornley} {et~al.}(2000){Thornley}, {Schreiber}, {Lutz}, {Genzel},
  {Spoon}, {Kunze}, \& {Sternberg}}]{thornley00}
{Thornley}, M.~D., {Schreiber}, N.~M.~F., {Lutz}, D., {et~al.} 2000, ApJ, 539,
  641

\bibitem[{{Tielens}(1993)}]{Tielens:93}
{Tielens}, A.~G.~G.~M. 1993, in Dust and Chemistry in Astronomy, ed. T.~J.
  {Millar} \& D.~A. {Williams}, 103

\bibitem[{{Tielens} {et~al.}(1993){Tielens}, {Meixner}, {van der Werf},
  {Bregman}, {Tauber}, {Stutzki}, \& {Rank}}]{Tielens:anatomyorionbar:93}
{Tielens}, A.~G.~G.~M., {Meixner}, M.~M., {van der Werf}, P.~P., {et~al.} 1993,
  "Science", 262, 86

\bibitem[{{Turner} {et~al.}(1998){Turner}, {Ho}, \& {Beck}}]{turner98}
{Turner}, J.~L., {Ho}, P.~T.~P., \& {Beck}, S.~C. 1998, AJ, 116, 1212

\bibitem[{{Vacca} \& {Conti}(1992)}]{vacca92}
{Vacca}, W.~D. \& {Conti}, P.~S. 1992, ApJ, 401, 543

\bibitem[{{Vacca} {et~al.}(2002){Vacca}, {Johnson}, \& {Conti}}]{vacca02}
{Vacca}, W.~D., {Johnson}, K.~E., \& {Conti}, P.~S. 2002, AJ, 123, 772

\bibitem[{{Vader} {et~al.}(1993){Vader}, {Frogel}, {Terndrup}, \&
  {Heisler}}]{vader93}
{Vader}, J.~P., {Frogel}, J.~A., {Terndrup}, D.~M., \& {Heisler}, C.~A. 1993,
  AJ, 106, 1743

\bibitem[{{Vanzi} \& {Rieke}(1997)}]{vanzi97}
{Vanzi}, L. \& {Rieke}, G.~H. 1997, ApJ, 479, 694

\bibitem[{{Vanzi} {et~al.}(1996){Vanzi}, {Rieke}, {Martin}, \&
  {Shields}}]{vanzi96}
{Vanzi}, L., {Rieke}, G.~H., {Martin}, C.~L., \& {Shields}, J.~C. 1996, ApJ,
  466, 150

\bibitem[{{Verma} {et~al.}(2003){Verma}, {Lutz}, {Sturm}, {Sternberg},
  {Genzel}, \& {Vacca}}]{verma03}
{Verma}, A., {Lutz}, D., {Sturm}, E., {et~al.} 2003, A\&A, 403, 829

\bibitem[{{Verstraete} {et~al.}(1996){Verstraete}, {Puget}, {Falgarone},
  {Drapatz}, {Wright}, \& {Timmermann}}]{Verstraete:m17:96}
{Verstraete}, L., {Puget}, J.~L., {Falgarone}, E., {et~al.} 1996, \aap, 315,
  L337

\bibitem[{{Viegas} {et~al.}(1999){Viegas}, {Contini}, \& {Contini}}]{viegas99}
{Viegas}, S.~M., {Contini}, M., \& {Contini}, T. 1999, A\&A, 347, 112

\bibitem[{{Walsh} \& {Roy}(1993)}]{walsh93}
{Walsh}, J.~R. \& {Roy}, J. 1993, MNRAS, 262, 27

\bibitem[{{Westera} {et~al.}(2004){Westera}, {Cuisinier}, {Telles}, \&
  {Kehrig}}]{westera04}
{Westera}, P., {Cuisinier}, F., {Telles}, E., \& {Kehrig}, C. 2004, A\&A, 423,
  133

\bibitem[{{Wood} \& {Churchwell}(1989)}]{wood89}
{Wood}, D.~O.~S. \& {Churchwell}, E. 1989, ApJS, 69, 831

\bibitem[{{Wynn-Williams} \& {Becklin}(1986)}]{wynn86}
{Wynn-Williams}, C.~G. \& {Becklin}, E.~E. 1986, ApJ, 308, 620

\bibitem[{{Young Owl} {et~al.}(2002){Young Owl}, {Meixner}, {Fong}, {Haas},
  {Rudolph}, \& {Tielens}}]{YoungOwl:02}
{Young Owl}, R.~C., {Meixner}, M.~M., {Fong}, D., {et~al.} 2002, ApJ, 578, 885

\end{thebibliography}

\end{document}